\documentclass[11pt]{article}
\usepackage{geometry} 
\usepackage[english]{babel}
\usepackage[english]{layout}
\usepackage{cite}

\usepackage{color}
\usepackage{mathpazo} 
\usepackage[scaled]{helvet} 
\usepackage{courier} 
\normalfont
\usepackage[T1]{fontenc}

\usepackage{amsfonts,latexsym,amsthm,amssymb}
\usepackage{amsmath,bbm}  

\def\fg{{\mathfrak g}}

\newcommand{\beq}{\begin{equation}}  
\newcommand{\eeq}{\end{equation}} 
\newcommand{\bea}{\begin{eqnarray}}  
\newcommand{\eea}{\end{eqnarray}} 
\newcommand{\beano}{\begin{eqnarray*}}  
\newcommand{\eeano}{\end{eqnarray*}} 
  
\newcommand{\mb}[1]{\quad\mbox{#1}\quad}

\usepackage{memhfixc}  
\usepackage{fancyhdr} 
\usepackage{sectsty}
\allsectionsfont{\sffamily\mdseries\upshape} 

\numberwithin{equation}{section}

\begin{document}

\title{\LARGE \textbf{Algebraic Bethe ansatz for the XXZ Heisenberg spin chain with triangular boundaries and the corresponding Gaudin model}}

\author{
\textsf{N. ~Manojlovi\'c}
\thanks{E-mail address: nmanoj@ualg.pt}
\textsf{ and I. ~Salom}
\thanks{E-mail address: isalom@ipb.ac.rs} \\
\\
\textit{$^{\ast}$Departamento de Matem\'atica, F. C. T.,
Universidade do Algarve} \\
\textit{Campus de Gambelas, PT-8005-139 Faro, Portugal}\\
\\
\textit{$^{\dag}$Institute of Physics, University of Belgrade}\\
\textit{P.O. Box 57, 11080 Belgrade, Serbia}\\
\\
}
\date{}


\maketitle
\thispagestyle{empty}
\begin{abstract}
The implementation of the algebraic Bethe ansatz for the XXZ Heisenberg spin chain 
in the case, when both reflection matrices have the upper-triangular form is analyzed. The general form of the Bethe vectors is studied. In the particular form, Bethe vectors admit the recurrent procedure, with an appropriate modification, used previously in the case of the XXX Heisenberg chain. As expected, these Bethe vectors yield the strikingly simple expression for the off-shell action of the transfer matrix of the chain as well as the spectrum of the transfer matrix and the corresponding Bethe equations. As in the XXX case, the so-called quasi-classical limit gives the off-shell action of the generating function of the corresponding trigonometric Gaudin Hamiltonians with boundary terms.

\end{abstract}

\clearpage
\newpage

\section{Introduction \label{sec: intro}}
The quantum inverse scattering method (QISM) is an approach to construct and solve quantum integrable systems \cite{TakhtajanFaddeev79, KulishSklyanin82, Faddeev98}. In the framework of the QISM the algebraic Bethe ansatz is a powerful algebraic approach, which yields the spectrum and corresponding eigenstates for the systems for which highest weight type representations are relevant, like for example quantum spin systems, Gaudin models, etc. In particular, the Heisenberg spin chain \cite{Heisenberg28}, with periodic boundary conditions, has been studied by the algebraic Bethe ansatz \cite{TakhtajanFaddeev79,Faddeev98}, including the question of completeness and simplicity of the spectrum \cite{Tarasov09}.

A way to introduce non-periodic boundary conditions compatible with the integrability of one-dimensional solvable quantum systems was developed in \cite{Sklyanin88}. The boundary conditions are expressed in the form of the left and right reflection matrices. The compatibility conditions between the bulk and the boundary of the system take the form of the so-called reflection equation, at the left site, and  the dual reflection equation, at the right site of the system. The matrix form of the exchange relations between the entries of the Sklyanin monodromy matrix is analogous to the reflection equation. Together with the dual reflection equation they yield the commutativity of the open transfer matrix \cite{Sklyanin88, FreidelMaillet91, FreidelMaillet91a}.

There is a renewed interest in applying the algebraic Bethe ansatz to the open XXX and XXZ chains with non-periodic boundary conditions compatible with the integrability of the systems \cite{Martins05, Eric13, Belliard13, CAMS, Lima13, Belliard15, BelliardPimenta15, AvanEtal15,Nepomechie15}. Other approaches include the Bethe ansatz based on the functional relation between the eigenvalues of the transfer matrix and the quantum determinant and the associated T-Q relation \cite{LucEricRafael07,China13, China15}, functional relations  for the eigenvalues of the transfer matrix based on fusion hierarchy \cite{Nepomechie04} and the Vertex-IRF correspondence \cite{Jimbo95, China03}. For a review of the coordinate Bethe ansatz for non-diagonal boundaries see \cite{Eric2013RMP}. For the latest results, as well as an excellent review, on the application of the separation of variables method on the 6-vertex model and the associate XXZ quantum chains see 
\cite{Maillet17}. However, we will focus on applying the algebraic Bethe ansatz to the 
XXZ Heisenberg spin chain in the case when system admits the so-called pseudo-vacuum, or the reference state. In his seminal work on boundary conditions in quantum integrable models Sklyanin has studied the XXZ spin chain with diagonal boundaries \cite{Sklyanin88}. As opposed to the case of the open XXX Heisenberg chain were both reflection matrices can be simultaneously brought to a triangular form by a single similarity transformation which leaves the R-matrix invariant and it is independent of the spectral parameter \cite{ Eric13,Belliard13,CAMS}, here the triangularity of the K-matrices has to be imposed by hand. The algebraic Bethe ansatz was applied to the
XXZ spin-$\textstyle{\frac{1}{2}}$ chain with upper triangular reflection matrices \cite{Lima13, Belliard15}. The spectrum and the corresponding Bethe equations were obtained \cite{Lima13} and 
the Bethe vectors were defined using a family of creations operators \cite{Belliard15}.

This work is centered on the study of the Bethe vectors which are fundamental in the implementation of the algebraic Bethe ansatz for the XXZ Heisenberg spin chain when the corresponding reflection matrices have the upper-triangular form. Seeking the Bethe vectors $\widetilde{\Psi} _M ( \mu_1 ,  \mu_2 ,  \dots ,  \mu_M )$ which would in the scaling limit coincide with the ones of the XXX Heisenberg chain \cite{CAMS}, we have also found certain identities yielding the general form of the Bethe vectors for a fixed $M$. The general form of Bethe vectors is given as a sum of a particular vector and the linear combination of lower order Bethe vectors. Due to certain identities this linear combination of lower order Bethe vectors corresponds the same eigenvalue as the particular vector. Although we have obtained explicitly the Bethe vectors $\widetilde{\Psi} _M ( \mu_1 ,  \mu_2 ,  \dots ,  \mu_M )$ for $M = 1, 2, 3, 4$,  unfortunately they do not admit a compact closed form for an arbitrary $M$. However, a detailed analysis yields a particular form of the Bethe vectors $\Psi _M ( \mu_1 ,  \mu_2 ,  \dots ,  \mu_M )$ which admits the recurrence formulas for the coefficient functions analogous to the once used in the study of the XXX Heisenberg chain \cite{CAMS}.  These Bethe vectors are defined explicitly, for an arbitrary natural number $M$, as some polynomial functions of the creation operators. Also, the off-shell action of the transfer matrix on these Bethe vectors is strikingly simple since it almost coincides with the corresponding action in the case when the two boundary matrices are diagonal. As expected, the off-shell action yields the spectrum of the transfer matrix and the corresponding Bethe equations.  To explore further these results we use the so-called quasi-classical limit and obtain the off-shell action of the generating function of the trigonometric Gaudin Hamiltonians with boundary terms, on the corresponding Bethe vectors. 

Originally in his approach, Gaudin defined these models as a quasi-classical limit of the integrable quantum chains \cite{Gaudin76,Gaudin83}. The Gaudin models were extended to any simple Lie algebra, with arbitrary irreducible representation at each site of the chain \cite{Gaudin83}. Sklyanin studied the rational $s\ell(2)$ model in the framework of the quantum inverse scattering method using the $s\ell(2)$ invariant classical r-matrix \cite{Sklyanin89}. A generalization of these results to all cases when skew-symmetric r-matrix satisfies the classical Yang-Baxter equation \cite{BelavinDrinfeld} was relatively straightforward \cite{SklyaninTakebe,Semenov97}. Therefore, considerable attention has been devoted to Gaudin models corresponding to the classical r-matrices of simple Lie algebras \cite{Jurco89, Jurco90, WagnerMacfarlane00} and Lie superalgebras \cite{BrzezinskiMacfarlane94, KulishManojlovic01, KulishManojlovic03, LimaUtiel01, KurakLima04}. 

Hikami showed how the quasi-classical expansion of the XXZ transfer matrix, calculated at the special values of the spectral parameter, yields the Gaudin Hamiltonians in the case when both reflection matrices are diagonal \cite{Hikami95}. Then the algebraic Bethe ansatz was applied to open Gaudin model in the context of the Vertex-IRF correspondence \cite{YangZhang12, YangZhangSasakic04, YangZhangSasakic05}. Also,  results were obtained for the open Gaudin models based on Lie superalgebras \cite{Lima09}. An approach to study the open Gaudin models based on the classical reflection equation \cite{Sklyanin87} and the non-unitary r-matrices \cite {Maillet85,Maillet86,Avan90} was developed, see \cite{Hikami95b, Skrypnyk09, Skrypnyk13, CAMRS, Skrypnyk16} and the references therein. For a review of the open Gaudin model see \cite{CAMN}.

In \cite{MS2} we have derived the generating function of the trigonometric Gaudin Hamiltonians with boundary terms following Sklyanin's approach for the periodic boundary conditions \cite{Sklyanin89, MNS}. Analogously to the rational case \cite{CAMRS, CAMS}, our derivation is based on the quasi-classical expansion of the linear combination of the transfer matrix of the XXZ Heisenberg chain and the central element, the so-called Sklyanin determinant. Here we use this result with the objective to derive the off-shell action of the generating function. As we will show below, the quasi-classical expansion of the Bethe vectors we have defined for the XXZ Heisenberg spin chain yields the Bethe vectors of the corresponding Gaudin model. The importance of these Bethe vectors stems from the striking simplicity of the off-shell action of the generating function of the trigonometric Gaudin Hamiltonians with boundary terms.

This paper is organized as follows. In Section \ref{sec: XXZ-chain} we review the suitable R-matrix as well as the Lax operator and the corresponding monodromy as the fundamental tools of the quantum inverse scattering method in the study of the inhomogeneous XXZ Heisenberg spin chain. The general solutions of the relevant reflection equation and the corresponding dual reflection equation are surveyed in Section \ref{sec: RE}. In Section \ref{sec: b-XXZ-chain} we briefly expose the Sklyanin approach to the inhomogeneous XXZ Heisenberg spin chain with non-periodic boundary conditions, in particular the derivation of the  relevant commutation relations. The implementation of the algebraic Bethe ansatz and most notably the study of the Bethe vectors, as one of the main results of the paper, are presented in Section \ref{sec: ABA-XXZchain}. The corresponding Gaudin model is studied through the quasi-classical limit in Section \ref{sec: Gm}. Our conclusions are presented in the Section \ref{sec: Conclu}.  In Appendix \ref{app: basic-def} are given some basic definitions for the convenience of the reader. The commutation relations relevant for the implementation of the algebraic Bethe ansatz for the XXZ Heisenberg chain are given in the appendix \ref{app:  commut-rel}. Finally, detailed presentation of the illustrative example of the Bethe vector $\widetilde{\Psi} _3 ( \mu_1 ,  \mu_2, \mu_3)$, including its general form and some important identities, are given in Appendix \ref{app: B-vec}.

\section{Inhomogeneous XXZ  Heisenberg spin chain \label{sec: XXZ-chain}}
The starting point in our study of the XXZ  Heisenberg spin chain is the R-matrix \cite{TakhtajanFaddeev79, KulishSklyanin82, Baxter72, Baxter82}
\begin{equation}
\label{XXZ-Rmatrix}
R (\lambda, \eta) = \left(\begin{array}{cccc}
\sinh(\lambda + \eta) & 0 & 0 & 0 \\[1ex]
0 & \sinh(\lambda) & \sinh(\eta) & 0 \\[1ex]
0 & \sinh(\eta) & \sinh(\lambda) & 0 \\[1ex]
0 & 0 & 0 & \sinh(\lambda + \eta) \end{array}\right) .
\end{equation}
This R-matrix satisfies the Yang-Baxter equation  \cite {Yang67, Baxter72, Baxter82, TakhtajanFaddeev79, KulishSklyanin82} in the space 
$\mathbb{C}^2 \otimes \mathbb{C}^2 \otimes \mathbb{C}^2$
\begin{equation}
\label{YBE}
R_{12} ( \lambda - \mu) R_{13} ( \lambda) R_{23} (\mu) = R_{23} (\mu ) R_{13} (\lambda ) R_{12} ( \lambda - \mu) ,
\end{equation}
and it also has other relevant properties such as

\begin{tabbing}{|l}
xxxxxxxxxxxxxxxxxxxxxxxxxx   \= xxxxxxxxxxxxxxxxxxxxxxxxxx     \kill
$U(1)$ symmetry \> $\left[ \sigma_1^3 + \sigma_2^3,  R_{12} ( \lambda )\right]=0$; \\
unitarity    \>   $R_{12} ( \lambda ) R_{21} ( -\lambda ) = \sinh (\eta - \lambda ) \sinh (\eta + \lambda )  \mathbbm{1}$;    \\
parity invariance    \> $R_{21} ( \lambda ) = R_{12} ( \lambda ) $;  \\
temporal invariance  \> $R_{12}^t ( \lambda) = R_{12} ( \lambda)$; \\
crossing symmetry  \>  $R ( \lambda) =   \mathcal{J} _1 R ^{t_2}( -\lambda - \eta ) {\mathcal{J}}_1,$
\end{tabbing}
where $t_2$ denotes the transpose in the second space and the two-by-two matrix $\mathcal{J}$ is proportional to the Pauli matrix $\sigma ^2$, i.e. $\mathcal{J} = \imath \sigma ^2$.

Here we study the inhomogeneous XXZ spin chain with $N$ sites, characterized by the local space 
$V_ m = \mathbb{C}^{2s+1}$ and inhomogeneous parameter $\alpha _m$. The Hilbert space of the system is
\begin{equation}
\label{H-space}
\mathcal{H} = \underset {m=1}{\overset {N}{\otimes}}  V_m = (\mathbb{C}^{2s+1} ) ^{\otimes N}.
\end{equation}
We introduce the Lax operator \cite{KulishResh83,Jimbo85,Jimbo86,FRT89,KulishSklyanin91,Zabrodin95,Anastasia07} as the following two-by-two matrix in the auxiliary space $V_0 = \mathbb{C}^2$,
\begin{equation}
\label{L-XXZ}
\mathbb{L}_{0m}(\lambda)  
= \frac{1}{\sinh (\lambda)} \left(\begin{array}{cc}  
\sinh \left( \lambda \mathbbm{1} _m + \eta S_m^{3} \right) & \sinh (\eta)  S_m^{-} \\[1ex] 
\sinh (\eta) S_m^{+} & \sinh \left( \lambda \mathbbm{1} _m - \eta S_m^{3} \right) 
\end{array}\right),
\end{equation}
the operators $S_m^{\alpha}$, with $\alpha = +,-, 3$ and $m= 1, 2 ,\dots , N$, are defined in the Appendix \ref{app: basic-def}. It obeys
\begin{equation}
\label{unit:Lax}
\mathbb{L}_{0m}(\lambda) \mathbb{L}_{0m}(\eta-\lambda) = \frac{\sinh \left( s_m \eta +\lambda \right) \sinh\left((s_m + 1) \eta - \lambda \right) }{\sinh (\lambda) \sinh (\eta - \lambda)}
\mathbbm{1} _0 \,,
\end{equation}
where $s_m$ is the value of spin in the space $V_m$. 

When the quantum space is also a spin $\frac12$ representation, the Lax operator becomes the $R$-matrix,
$$\mathbb{L}_{0m}(\lambda) = \frac{1}{\sinh (\lambda)} R_{0m}\left(\lambda - {\eta}/{2} \right). $$

Taking into account the commutation relations \eqref{crspin-s}, it is straightforward to check that the Lax operator satisfies the RLL-relations
\begin{equation}
\label{RLL}
R_{00'} ( \lambda - \mu) \mathbb{L}_{0m}( \lambda - \alpha _m ) \mathbb{L}_{0'm}( \mu - \alpha _m ) =  \mathbb{L}_{0'm}( \mu  - \alpha _m ) \mathbb{L}_{0m}( \lambda  - \alpha _m )R_{00'} ( \lambda - \mu).
\end{equation}
The so-called monodromy matrix
\begin{equation}
\label{monodromy-T}
T(\lambda ) = \mathbb{L}_{0N} ( \lambda - \alpha _N) \cdots \mathbb{L}_{01} ( \lambda - \alpha _1)
\end{equation}
is used to describe the system. For simplicity we have omitted the dependence on the quasi-classical parameter $\eta$ and the inhomogeneous parameters $\{ \alpha _j , j = 1 , \ldots , N \}$. Notice that $T(\lambda)$ is a two-by-two matrix acting in the auxiliary space $V_0 = \mathbb{C}^2$, whose entries are operators acting in $\mathcal{H}$
\begin{equation}
\label{T-mat}
T(\lambda ) = \left(\begin{array}{cc} 
A(\lambda ) &  B(\lambda ) \\ 
C(\lambda ) &  D(\lambda ) \end{array}\right) .
\end{equation}
From RLL-relations \eqref{RLL} it follows that the monodromy matrix satisfies the RTT-relations
\begin{equation}
\label{RTT}
R_{00'} ( \lambda - \mu) {T}_{0} (\lambda ) {T}_ {0'}(\mu ) =  {T} _ {0'}(\mu ){T}_ {0}(\lambda ) R_{00'} ( \lambda - \mu).
\end{equation}

To construct integrable spin chains with non-periodic boundary condition, we will follow Sklyanin's approach \cite{Sklyanin88}. Accordingly, before defining the essential operators and corresponding algebraic structure, in the next section we will introduce the relevant boundary K-matrices.

\section{Reflection equation \label{sec: RE}}
A way to introduce non-periodic boundary conditions which are compatible with the integrability of the bulk model, was developed in \cite{Sklyanin88}. Boundary conditions on the left and right sites of the chain are encoded in the left and right reflection matrices $K^-$ and $K^+$. The compatibility condition between the bulk and the boundary of the system takes the form of the so-called reflection equation. It is written in the following form for the left reflection matrix acting on the space $\mathbb{C}^2$ at the first site $K^-(\lambda) \in \mathrm{End} (\mathbb{C}^2)$
\begin{equation}
\label{RE}
R_{12}(\lambda - \mu) K^-_1(\lambda) R_{21}(\lambda + \mu) K^-_2(\mu)=
K^-_2(\mu) R_{12}(\lambda + \mu) K^-_1(\lambda) R_{21}(\lambda - \mu) .
\end{equation}

Due to the properties of the R-matrix \eqref{XXZ-Rmatrix} the dual reflection equation can be presented in the following form
\begin{equation}
\label{dRE}
R_{12}( \mu-\lambda )K_1^{+}(\lambda) R_{21}(-\lambda - \mu - 2\eta)  K_2^{+}(\mu)=
K_2^{+}(\mu) R_{12}(-\lambda -\mu-2\eta) K_1^{+}(\lambda) R_{21}(\mu-\lambda) .
\end{equation}
One can then verify that the mapping
\begin{equation}
\label{bijectionKpl}
K^+(\lambda)= K^{-}(- \lambda -\eta)
\end{equation}
is a bijection between solutions of the reflection equation and the dual reflection equation. After substitution of \eqref{bijectionKpl} into the dual reflection equation \eqref{dRE} one gets the reflection equation \eqref{RE} with shifted arguments.

The general, spectral parameter dependent, solutions of the reflection equation \eqref{RE} and the dual reflection equation \eqref{dRE} can be written as follows \cite{VegaGonzalez, Zamolodchikov94,Zamolodchikov94b}
\begin{align}
\label{K-minus}
K ^{-}(\lambda) &=  \left(\begin{array}{cc}
\kappa ^{-} \sinh (\xi ^{-} + \lambda) & \psi ^{-} \sinh (2 \lambda) \\ \phi ^{-} \sinh (2 \lambda) & \kappa ^{-}\sinh (\xi ^{-} - \lambda) \end{array}\right) , \\[2ex]
\label{K-plus}
K ^{+}(\lambda) &=  
\left(\begin{array}{cc}
\kappa ^{+} \sinh ( \xi ^{+} - \lambda - \eta) & - \psi ^{+} \sinh \left( 2 (\lambda + \eta) \right) \\ 
- \phi ^{+} \sinh \left( 2 (\lambda + \eta) \right) & \kappa ^{+}\sinh (\xi ^{+} + \lambda + \eta)
\end{array}\right) .
\end{align}
Due to the fact that the reflection matrices $K ^{\mp}(\lambda)$ are defined up to multiplicative constants the values of parameters $\kappa ^{\mp}$ are not essential, as long as they are different from zero. Therefore they could be set to be one without any loss of generality. In particular, this will be evident throughout the Sections \ref{sec: ABA-XXZchain} and \ref{sec: Gm}. However, for completeness, we will keep them in our presentation. 

Although the R-matrix \eqref{XXZ-Rmatrix} has the $U(1)$ symmetry the reflection matrices $K ^{\mp}(\lambda)$ \eqref{K-minus} and \eqref{K-plus} cannot be brought to the upper triangular form by the symmetry transformations like in the case of the XXX Heisenberg spin chain \cite{Eric13,CAMS}. Therefore, as we will see in the Section \ref{sec: ABA-XXZchain}, triangularity of the reflections matrices has to be imposed as extra conditions on the parameters of the reflection matrices.

\section{Inhomogeneous XXZ Heisenberg spin chain with boundary terms \label{sec: b-XXZ-chain}}
In order to develop the formalism necessary to describe an integrable spin chain with non-periodic boundary condition, we use the Sklyanin approach \cite{Sklyanin88}. The main tool in this framework is the corresponding monodromy matrix  
\begin{equation}
\label{cal-T}
\mathcal{T}_0(\lambda)= T_0(\lambda) K _0^{-}(\lambda) \widetilde T_0(\lambda),
\end{equation}
it consists of the matrix $T(\lambda)$ \eqref{monodromy-T},  a reflection matrix $K ^{-}(\lambda)$ \eqref{K-minus} and the matrix
\begin{equation}
\label{tilde-T}
\begin{split}
\widetilde {T}_0(\lambda)= 
\left(\begin{array}{cc} 
\widetilde{A}(\lambda ) &  \widetilde{B}(\lambda ) \\ 
\widetilde{C}(\lambda ) &  \widetilde{D}(\lambda ) 
\end{array}\right) = \mathbb{L}_{01}(\lambda + \alpha _1 + \eta) \cdots \mathbb{L}_{0N} (\lambda + \alpha _N + \eta) .
\end{split}
\end{equation}
It is important to notice that the identity \eqref{unit:Lax}
can be rewritten in the form
\begin{equation}
\label{unit:LaxII}
\mathbb{L}_{0m}(\lambda - \alpha _m) \mathbb{L}_{0m}(\eta - \lambda + \alpha _m ) 
= \Big(\frac{\sinh \left( \lambda - \alpha _m + s_m \eta \right) \sinh\left( - \lambda + \alpha _m + (s_m + 1) \eta\right) }{\sinh (\lambda - \alpha _m) \sinh (- \lambda + \alpha _m + \eta )}  \Big) 
\mathbbm{1} _0 \, .
\end{equation}
It follows from the equation above and the RLL-relations \eqref{RLL} that the RTT-relations \eqref{RTT} can be recast as follows
\begin{align}
\label{tTRT}
\widetilde{T} _ {0'}(\mu ) R_{00'} ( \lambda + \mu) T _{0} (\lambda )  &= T _ {0}(\lambda ) R_{00'} ( \lambda + \mu) \widetilde{T} _ {0'}(\mu ) , \\
\label{tTtTR}
\widetilde{T} _{0} (\lambda ) \widetilde{T} _ {0'}(\mu ) R_{00'} (\mu - \lambda)   &= R_{00'} (\mu - \lambda) \widetilde{T} _ {0'}(\mu ) \widetilde{T} _ {0}(\lambda ) .
\end{align}
Using the RTT-relations \eqref{RTT}, \eqref{tTRT}, \eqref{tTtTR} and the reflection equation \eqref{RE} it is straightforward to show that the exchange relations of the monodromy matrix $\mathcal{T}(\lambda)$  in $V_0\otimes V_{0'}$ are
\begin{equation}
\label{RE-algebra}
R _{00'}(\lambda - \mu) \mathcal{T}_{0} (\lambda) R _{0'0} (\lambda + \mu) \mathcal{T} _{0'} (\mu) = 
\mathcal{T}_{0'}(\mu) R _{00^{\prime}} (\lambda + \mu) \mathcal{T}_{0} (\lambda) 
R _{0'0} (\lambda - \mu) ,
\end{equation}
using the notation of \cite{Sklyanin88}. From the above equation  we can read off the commutation relations of the entries of the monodromy matrix
\begin{equation}
\label{calT-mat}
\mathcal{T}(\lambda) = \left(\begin{array}{cc}
\mathcal{A} (\lambda) & \mathcal{B} (\lambda)  \\ 
\mathcal{C} (\lambda) & \mathcal{D} (\lambda)  \end{array} \right) .
\end{equation}
Following Sklyanin \cite{Sklyanin88}, as in the case of the XXX Heisenberg spin chain \cite{Eric13,CAMS}, we introduce the operator
\begin{equation}
\label{D-hat}
\widehat{\mathcal{D}} (\lambda) = \mathcal{D} (\lambda) - \frac{\sinh (\eta)}{\sinh (2\lambda + \eta)} \mathcal{A} (\lambda) .
\end{equation}
For convenience, the commutation relations relevant for the implementation of the algebraic Bethe ansatz for the XXZ Heisenberg chain are given in the appendix \ref{app:  commut-rel}.

The exchange relations \eqref{RE-algebra} admit a central element,  the so-called Sklyanin determinant,
\begin{equation}
\label{Delta-T-cal}
\Delta \left[\mathcal{T}(\lambda)\right] = \mathrm{tr}_{00'} P^{-}_{00'} \mathcal{T}_{0}(\lambda-\eta/2) R_{00'} (2\lambda) \mathcal{T}_{0'}(\lambda+\eta/2). 
\end{equation}
Analogously to the XXX Heisenberg spin chain \cite{CAMS}, the element $\Delta \left[\mathcal{T}(\lambda)\right]$ can be expressed in form
\begin{equation}
\label{Del-calT}
\Delta \left[\mathcal{T}(\lambda)\right] =  \sinh (2 \lambda) \widehat{\mathcal{D}}  (\lambda - \eta/2) \mathcal{A} (\lambda + \eta/2 )  - \sinh (2 \lambda + \eta ) \mathcal{B} (\lambda - \eta/2)  \mathcal{C} (\lambda + \eta/2 ).
\end{equation}

The open chain transfer matrix is given by the trace of the monodromy $\mathcal{T}(\lambda)$ over the auxiliary space $V_0$ with an extra reflection matrix $K^+(\lambda)$ \cite{Sklyanin88},
\begin{equation}
\label{open-t}
t (\lambda) = \mathrm{tr}_0 \left( K^+_0(\lambda) \mathcal{T}_0(\lambda) \right).
\end{equation}
The reflection matrix $K^+(\lambda)$ \eqref{K-plus} is the corresponding solution of the dual reflection equation \eqref{dRE}. The commutativity of the transfer matrix for different values of the spectral parameter
\begin{equation}
\label{open-tt}
[t (\lambda) , t (\mu)] = 0,
\end{equation}
is guaranteed by the dual reflection equation \eqref{dRE} and the exchange relations \eqref{RE-algebra} of the monodromy matrix $\mathcal{T}(\lambda)$ \cite{Sklyanin88}.

\section{Algebraic Bethe Ansatz \label{sec: ABA-XXZchain}}
In this section, we study the implementation of the algebraic Bethe ansatz for the XXZ Heisenberg spin chain 
when both reflection matrices $K^{\mp}(\lambda)$ are upper triangular. As opposed to the case of the XXX Heisenberg spin chain where the general reflection matrices could be put into the upper triangular form without any loss of generality \cite{Eric13, CAMS}, here the triangularity of the reflection matrices has to be imposed as extra conditions on the parameters of the reflection matrices $K^{\mp}(\lambda)$ \eqref{K-minus} and \eqref{K-plus}. Our aim is to obtain the Bethe vectors whose scaling limit corresponds to the ones of the XXX Heisenberg  chain \cite{CAMS}.

As our starting point in the implementation of the algebraic Bethe ansatz, we observe that in every $V_ m = \mathbb{C}^{2s+1}$ there exists a vector $\omega_m \in V_ m$ such that
\begin{equation}
\label{S-on-om}
S^3_m \omega _m = s_m \omega _m  \quad \text{and}  \quad S^+_m \omega _m = 0 .
\end{equation}
We define a vector $\Omega _+$ to be
\begin{equation}
\label{Omega+}
\Omega _+ = \omega _1 \otimes \cdots \otimes \omega _N \in \mathcal{H}.
\end{equation} 
From the definitions \eqref{L-XXZ}, \eqref{monodromy-T} and \eqref{S-on-om} it is straightforward to obtain the action of the entries of the monodromy matrix $T(\lambda )$ \eqref{monodromy-T} on the vector $\Omega_+$ 
\begin{eqnarray}
\label{AonOm}
A(\lambda) \Omega_+  &=&  a (\lambda) \Omega_+ , \quad \text{with}  \quad  a(\lambda) = \prod _{m=1}^N \frac{\sinh ( \lambda - \alpha _m + \eta s_m)}{\sinh (\lambda - \alpha _m)} , \\
\label{DonOm}
D(\lambda) \Omega_+  &=&  d (\lambda) \Omega_+ , \quad \text{with}  \quad  d(\lambda) = \prod _{m=1}^N \frac{\sinh (\lambda - \alpha _m - \eta s_m)}{\sinh (\lambda - \alpha _m)} , \\
\label{ConOm}
C(\lambda) \Omega_+  &=& 0.                                              
\end{eqnarray}
Analogously, from the definitions \eqref{L-XXZ}, \eqref{tilde-T} and \eqref{S-on-om} it is straightforward to obtain the action of the entries of the monodromy matrix $\widetilde {T}(\lambda)$ \eqref{tilde-T} on the vector $\Omega_+$ 
\begin{eqnarray}
\label{AtilonOm}
\widetilde{A}(\lambda ) \Omega_+ &=& \widetilde{a} (\lambda) \Omega_+ , \quad \text{with}  \quad  \widetilde{a}(\lambda) = \prod _{m=1}^N \frac{\sinh (\lambda + \alpha _m + \eta (1 + s_m) )}{\sinh (\lambda + \alpha _m + \eta)} , \\
\label{DtilonOm}    
\widetilde{D}(\lambda) \Omega_+  &=&  \widetilde{d}(\lambda) \Omega_+ , \quad \text{with}  \quad  \widetilde{d}(\lambda) = \prod _{m=1}^N \frac{\sinh ( \lambda + \alpha _m + \eta ( 1 - s_m) )}{\sinh (\lambda + \alpha _m + \eta)} , \\
\label{CtilonOm}
\widetilde{C}(\lambda) \Omega_+  &=& 0 .  
\end{eqnarray}
Since the left reflection matrix cannot be brought to the upper triangular form by the $U(1)$ symmetry transformations we have to impose  an extra condition on the parameters of $K ^{-}(\lambda)$. By setting $$\phi ^{-}=0$$ in \eqref{K-minus} the reflection matrix $K ^{-}(\lambda)$ becomes upper triangular and according to definition of the Sklyanin monodromy matrix \eqref{cal-T} we have
\begin{equation}
\label{calT-tri}
\mathcal{T}(\lambda) = 
\left(\begin{array}{cc} 
A(\lambda ) &  B(\lambda ) \\ 
C(\lambda ) &  D(\lambda ) 
\end{array}\right) 
\left(\begin{array}{cc}
\kappa ^{-} \sinh (\xi ^{-} + \lambda) & \psi ^{-} \sinh (2 \lambda) \\ 
0 & \kappa ^{-}\sinh (\xi ^{-} - \lambda) 
\end{array}\right)
\left(\begin{array}{cc} 
\widetilde{A}(\lambda ) &  \widetilde{B}(\lambda ) \\ 
\widetilde{C}(\lambda ) &  \widetilde{D}(\lambda ) 
\end{array}\right)  .
\end{equation}
From the above equation, using the relations which follow from \eqref{tTRT}  we obtain
\begin{align}
\label{CalA}
\mathcal{A} (\lambda) &= \kappa ^{-} \sinh (\xi ^{-} + \lambda) \ A(\lambda) \widetilde{A}(\lambda) \notag \\
&+ \left( \psi ^{-} \sinh (2 \lambda) \, A(\lambda) + \kappa ^{-} \sinh (\xi ^{-} - \lambda) \, B(\lambda) \right)
\widetilde{C}(\lambda) \\
\label{CalD}
\mathcal{D} (\lambda) &=  \kappa ^{-} \sinh (\xi ^{-} + \lambda) \left( \widetilde{B}(\lambda) C(\lambda ) 
- \frac{\sinh (\eta)}{\sinh (2\lambda + \eta)} \left( D(\lambda ) \widetilde{D}(\lambda) - \widetilde{A}(\lambda ) A(\lambda ) \right) \right) \notag \\
&+ \left(  \psi ^{-} \sinh (2 \lambda) \ C (\lambda) + \kappa ^{-}\sinh (\xi ^{-} - \lambda) \ D(\lambda) \right) \widetilde{D}(\lambda) \\[1ex]
\label{CalB} 
\mathcal{B} (\lambda) &= \kappa ^{-} \sinh (\xi ^{-} + \lambda) \left( \frac{\sinh (2\lambda)}{\sinh (2\lambda + \eta)} \ \widetilde{B}(\lambda ) A(\lambda) - \frac{\sinh (\eta)}{\sinh (2\lambda + \eta)} \ B(\lambda ) \widetilde{D}(\lambda ) \right) \notag \\
&+ \left( \psi ^{-} \sinh (2 \lambda) \ A (\lambda) + \kappa ^{-}\sinh (\xi ^{-} - \lambda) \ B(\lambda) \right) \widetilde{D}(\lambda) \\[1ex]
\label{CalC}
\mathcal{C} (\lambda)&=  \kappa ^{-} \sinh (\xi ^{-} + \lambda) \ C(\lambda) \widetilde{A}(\lambda) + \left( \psi ^{-} \sinh (2 \lambda) \ C(\lambda) + \kappa ^{-}\sinh (\xi ^{-} - \lambda) \ D(\lambda) \right) \widetilde{C}(\lambda)  .
\end{align}
The action of the entries of the Sklyanin monodromy matrix on the vector $\Omega _+$ follows from the above relations \eqref{CalA}-\eqref{CalC} and the formulae \eqref{AonOm}-\eqref{ConOm} and \eqref{AtilonOm}-\eqref{CtilonOm} 
\begin{eqnarray}
\label{CalConOm}
\mathcal{C} (\lambda) \Omega_+ &=& 0, \\
\label{CalAonOm}
\mathcal{A} (\lambda) \Omega_+ &=& \alpha (\lambda) \Omega_+ , \quad \text{with}  \quad \alpha (\lambda) =  \kappa ^{-} \sinh (\xi ^{-} + \lambda) \ a(\lambda) \widetilde{a}(\lambda) , \\
\label{CalDonOm}
\mathcal{D} (\lambda) \Omega_+ &=& \delta (\lambda) \Omega_+ , \quad \text{with}  \\
\delta (\lambda) &=& \kappa ^{-} \left( \sinh (\xi ^{-} - \lambda) - \sinh (\xi ^{-} + \lambda) \frac{\sinh (\eta)}{\sinh (2\lambda + \eta)}    \right) d(\lambda) \widetilde{d}(\lambda) \notag \\ 
&+& \kappa ^{-} \sinh (\xi ^{-} + \lambda)  \frac{\sinh (\eta)}{\sinh (2\lambda + \eta)}  \ a(\lambda) \widetilde{a}(\lambda) .
\notag
\end{eqnarray}
In what follows we will also use the fact that $\Omega_+$ is an eigenvector of the operator $\widehat{\mathcal{D}} (\lambda)$ \eqref{D-hat} 
\begin{equation}
\label{DhatOm}
\widehat{\mathcal{D}} (\lambda) \Omega_+ = \widehat{\delta} (\lambda) \Omega_+ , \quad \text{with}  \quad \widehat{\delta} (\lambda) = \delta (\lambda) - \frac{\sinh (\eta)}{\sinh (2\lambda + \eta)} \ \alpha (\lambda) ,
\end{equation}
or explicitly 
\begin{equation}
\label{deltahat}
\widehat{\delta }(\lambda) = \kappa ^{-} \left( \sinh (\xi ^{-} - \lambda) - \sinh (\xi ^{-} + \lambda) \frac{\sinh (\eta)}{\sinh (2\lambda + \eta)}    \right) d(\lambda) \widetilde{d}(\lambda) . 
\end{equation}
The transfer matrix of the inhomogeneous XXZ chain 
\begin{equation}
\label{open-t}
t _0(\lambda) = \mathrm{tr}_0 \left( K^+(\lambda) \mathcal{T}(\lambda) \right) ,
\end{equation}
with the triangular K-matrix
\begin{equation}
K ^{+}(\lambda) = 
\left(\begin{array}{cc}
\kappa ^{+} \sinh ( \xi ^{+} - \lambda - \eta) & - \psi ^{+} \sinh \left( 2 (\lambda + \eta) \right) \\ 
 0 & \kappa ^{+}\sinh (\xi ^{+} + \lambda + \eta)
\end{array}\right) ,
\end{equation}
i.e. the matrix $K ^{+}(\lambda) = K^{-}(- \lambda -\eta)$ were we have set
$$
\phi ^{+} = 0 ,
$$
can be expressed using Sklyanin's $\widehat{\mathcal{D}} (\lambda)$ operator \eqref{D-hat} 
\begin{equation}
\label{transfer-matrix}
t(\lambda) = \kappa _1 (\lambda) \ \mathcal{A} (\lambda) + \kappa _2 (\lambda) \ \widehat{\mathcal{D}} (\lambda) + \kappa _{12} (\lambda) \ \mathcal{C} (\lambda),
\end{equation}
with
\begin{align}
\label{k-s}
\kappa _1 (\lambda) &=  \kappa ^{+} \sinh ( \xi ^{+} - \lambda) \ \frac{\sinh(2(\lambda + \eta))}{\sinh (2 \lambda + \eta)} , \quad
\kappa _2 (\lambda) = \kappa ^{+} \sinh (\xi ^+ + \lambda + \eta) , \notag \\[1ex]
\kappa _{12} (\lambda) &= - \psi^+  \sinh (2 (\lambda + \eta)) .
\end{align}

Evidently, due to \eqref{CalConOm}-\eqref{deltahat},  the vector $\Omega _+$ \eqref{Omega+} is an eigenvector  of the transfer matrix
\begin{equation}
\label{t-on-Om+}
t(\lambda) \Omega _+ = \left( \kappa _1 (\lambda) \alpha (\lambda) + \kappa _2 (\lambda) \widehat{\delta} (\lambda) \right) \Omega _+ = \Lambda _0 (\lambda) \Omega _+ . 
\end{equation}
For simplicity we have suppressed the dependence of the eigenvalue $\Lambda _0 (\lambda)$ on the boundary parameters $\kappa ^{+}$, $\xi ^+$ and $\psi^+$ as well as the quasi-classical parameter $\eta$.

Let us consider  
\begin{equation}
\label{Psi1-til}
\widetilde{\Psi} _1 (\mu) = \mathcal{B} (\mu) \Omega _+ - \frac{\psi^+}{\kappa ^+} 
\left(  \frac{ \sinh (2\mu)}{\sinh (2 \mu  + \eta)}  \cosh (\xi ^+ - \mu) \ \alpha (\mu ) -  \cosh ( \xi ^+ + \mu + \eta) \ \widehat{ \delta}  (\mu) \right) \Omega _+ .
\end{equation}
A straightforward calculation, using the relations \eqref{comm-relAB}, \eqref{comm-rel-hDB} and \eqref{comm-relCB}, shows that the off-shell action of the transfer matrix \eqref{transfer-matrix} on $\widetilde{\Psi } _1 (\mu)$ is given by
\begin{equation}
\label{t-on-Psi1-til}
t(\lambda) \widetilde{\Psi }_1 (\mu) = \Lambda _1 (\lambda , \mu ) \widetilde{\Psi} _1 (\mu) + \kappa ^+ \sinh ( \xi^+ - \mu ) \frac{\sinh (\eta) \sinh ( 2 (\lambda + \eta) )}{\sinh (\lambda - \mu) \sinh(\lambda + \mu + \eta)} F_1 (\mu) \widetilde{\Psi} _1 (\lambda) ,
\end{equation} 
where the eigenvalue $\Lambda _1 (\lambda , \mu )$  
is given by
\begin{equation}
\label{Lambda1}
\!\!\!\!\!\!\!\!\!\!
\Lambda _1 (\lambda , \mu ) =  \kappa _1 (\lambda) \frac{\sinh (\lambda + \mu) \sinh (\lambda - \mu - \eta)}{\sinh (\lambda - \mu) \sinh(\lambda + \mu + \eta)} \alpha (\lambda) + \kappa _2 (\lambda) \frac{\sinh (\lambda - \mu + \eta) \sinh(\lambda + \mu + 2 \eta)}{\sinh(\lambda - \mu) \sinh(\lambda + \mu + \eta)} \widehat{\delta} (\lambda) .
\end{equation}
Evidently $\Lambda _1 (\lambda , \mu )$ depends also on boundary parameters $\kappa ^+$, $\xi ^+$  and the quasi-classical parameter $\eta$, but these parameters are omitted in order to simplify the formulae. The unwanted term on the right hand side \eqref{t-on-Psi1-til} is annihilated by the Bethe equation 
\begin{equation}
\label{F1}
F_1 (\mu) = \frac{ \sinh (2\mu)}{\sinh (2 \mu  + \eta)} \ \alpha (\mu ) - \frac{\sinh (\xi^+ + \mu + \eta)}{\sinh (\xi^+ - \mu)} \ \widehat{ \delta}  (\mu) = 0,
\end{equation}
or equivalently,
\begin{equation}
\label{Bethe Eq-1}
\frac{\alpha (\mu)}{\widehat{ \delta} (\mu)} = \frac{\sinh (2 (\mu  + \eta)) \ \kappa _2 (\mu)}{\sinh (2 \mu) \ \kappa _1 (\mu)} = \frac{\sinh(2 \mu  + \eta) \sinh(\xi^+ + \mu + \eta)}{\sinh (2 \mu) \ \sinh (\xi^+ - \mu)} .
\end{equation}
Thus we have shown that $\widetilde{\Psi} _1 (\mu)$ \eqref{Psi1} is a Bethe vector of the transfer matrix \eqref{transfer-matrix}. Moreover, the vector $\widetilde{\Psi} _1 (\mu)$ in the scaling limit
yields the corresponding Bethe vector of the XXX Heisenberg spin chain \cite{CAMS} and it was this connection that led us to this particular form of the Bethe vector. However, it is important to note that this is not the only possible form of the Bethe vector. Namely, we notice the following important identity 
\begin{equation}
\label{identity-Psi1}
\Lambda _0 (\lambda) - \Lambda _1 (\lambda , \mu ) = \kappa ^+ \sinh ( \xi^+ - \lambda ) \frac{\sinh (\eta) \sinh ( 2 (\lambda + \eta) )}{\sinh (\lambda - \mu) \sinh(\lambda + \mu + \eta)} F_1 (\lambda) .
\end{equation}
It follows that  $\widetilde{\Psi }_1 (\mu)$ \eqref{Psi1-til} can be generalized by adding a term proportional to 
$F_1(\mu )$
\begin{equation}
\label{Psi1-tilde}
\widetilde{\Psi}_1 (\mu, C_1) = \widetilde{\Psi}_1 (\mu) + C _1 \ \frac{\psi^+}{\kappa ^+} \ \sinh ( \xi^+ - \mu ) F_1(\mu )\Omega _+ ,
\end{equation}
where $C _1$ is independent of $\mu$. A direct consequence of the above identity is the off-shell action of the transfer matrix on $\widetilde{\Psi}_1 (\mu, C_1)$, 
\begin{equation}
\label{t-on-Psi1-tilde}
t(\lambda) \widetilde{\Psi}_1 (\mu, C_1) = \Lambda _1 (\lambda , \mu ) \widetilde{\Psi}_1 (\mu, C_1) + \kappa ^+ \sinh ( \xi^+ - \mu ) \frac{\sinh (\eta) \sinh ( 2 (\lambda + \eta) )}{\sinh (\lambda - \mu) \sinh(\lambda + \mu + \eta)} F_1 (\mu) \widetilde{\Psi}_1 (\lambda, C_1) . 
\end{equation}
Therefore $\widetilde{\Psi}_1 (\mu, C_1)$ \eqref{Psi1-tilde} can be considered as the general form of the Bethe vector of the transfer matrix \eqref{transfer-matrix} corresponding to the eigenvalue $\Lambda _1 (\lambda , \mu )$ \eqref{Lambda1}. 

By setting $C _1 = 1$ in \eqref{Psi1-tilde} we obtain another particular solution for the Bethe vector, that will turn out to be more suitable for the recurrence procedure
\begin{equation}
\label{Psi1}
 \Psi_1 (\mu) = \widetilde{\Psi}_1 (\mu, C _1 = 1) = \mathcal{B} (\mu) \Omega _+ + b_ 1 (\mu) \Omega _+ ,
\end{equation}
where $b_ 1 (\mu)$ is given by
\begin{equation}
\label{b1}
b_ 1 (\mu) = \left( - \frac{\psi^+}{\kappa ^+} \right) \left(  \frac{ \sinh (2\mu)}{\sinh (2 \mu  + \eta)}  e^{-(\xi ^+ - \mu)} \ \alpha (\mu ) -  e^{-( \xi ^+ + \mu + \eta)} \ \widehat{ \delta}  (\mu) \right) .
\end{equation}

We seek the Bethe vector $\widetilde{\Psi} _2 ( \mu_1 ,  \mu_2)$ in the form 
\begin{equation}
\label{Psi2-til}
\widetilde{\Psi }_2 ( \mu_1 ,  \mu_2) = \mathcal{B} (\mu_1) \mathcal{B} (\mu_2) \Omega _+ +  \widetilde{b}^{(1)}_2(\mu_2 ; \mu_1)  \mathcal{B} (\mu_1) \Omega _+  +  \widetilde{b}^{(1)}_2(\mu_1; \mu_2 ) \mathcal{B} (\mu_2) \Omega _+ +  \widetilde{b}^{(2)}_2 ( \mu_1 ,  \mu_2) \Omega _+ .
\end{equation}
One possible choice of the coefficient functions $\widetilde{b}^{(1)}_2(\mu_1; \mu_2 )$ and $\widetilde{b}^{(2)}_2( \mu_1 ,  \mu_2)$ is given by
\begin{align}
\label{1b2-tilde}
\widetilde{b}^{(1)}_2 (\mu_1; \mu_2 ) &= \left( - \frac{\psi^+}{\kappa ^+} \right)
 \left( \frac{ \sinh (2 \mu _1)}{\sinh (2 \mu _1  + \eta)}
\frac{\sinh (\mu _1 + \mu _2) \sinh (\mu _1 - \mu _2 - \eta )}{\sinh (\mu _1 - \mu _2) \sinh (\mu _1 + \mu _2 + \eta )} \ \cosh (\xi ^+ - \mu _1) \ \alpha (\mu _1) \right. \notag \\[1ex]
&\left. - \frac{\sinh (\mu _1 - \mu_2 + \eta) \sinh (\mu _1 + \mu_2 + 2 \eta)}{\sinh (\mu _1 - \mu _2) \sinh (\mu _1 + \mu_2 + \eta)} \ \cosh (\xi ^+ + \mu _1 + \eta)  \ \widehat{ \delta}  (\mu _1) \right) , 
\end{align}
and
\begin{align}
\label{2b2-tilde}
\widetilde{b}^{(2)}_2 ( \mu_1 ,  \mu_2) &=  \left( \frac{\psi^+}{\kappa ^+} \right) ^2 \left( \frac{\sinh (\mu _1 + \mu_2)}{\sinh (\mu _1 + \mu _2 + \eta)} \, \frac{\sinh (2 \mu _1)}{\sinh (2 \mu _1 + \eta)} \, \frac{\sinh (2 \mu _2)}{\sinh (2 \mu _2 + \eta)} \, \cosh (2 \xi ^+ - \mu _1 - \mu _2 + \eta ) \times \right. \notag \\[1ex]
&  \times \alpha (\mu _1) \alpha (\mu _2) - \frac{\sinh (2 \mu _1)}{\sinh (2 \mu _1 + \eta)} \, \frac{\sinh (\mu_1 - \mu_2 - \eta)}{\sinh (\mu_1 - \mu_2)}  \, \cosh (2 \xi ^+ - \mu _1 + \mu _2 + 2 \eta ) \ \alpha (\mu _1)   \widehat{ \delta}  (\mu _2) \notag \\[1ex]
&- \frac{\sinh (\mu_2 - \mu_1 - \eta)}{\sinh (\mu_2 - \mu_1)}  \, \frac{\sinh (2 \mu _2)}{\sinh (2 \mu _2 + \eta)} \, \cosh (2 \xi ^+ + \mu _1 - \mu _2 + 2 \eta ) \ \widehat{ \delta} (\mu _1) \alpha (\mu _2) \notag \\[1ex]
&\left. + \frac{\sinh (\mu_1 + \mu_2 + 2 \eta)}{\sinh (\mu_1 + \mu_2 + \eta)} \, \cosh (2 \xi ^+ + \mu _1 + \mu _2 + 3 \eta ) \ \widehat{ \delta} (\mu _1) \widehat{ \delta} (\mu _2) \right) .
\end{align}
Due to the fact that the operators $\mathcal{B} (\mu_1)$ and $\mathcal{B} (\mu_2)$ commute \eqref{comm-relBB+CC} and that $\widetilde{b}^{(2)}_2 ( \mu_1 ,  \mu_2) = \widetilde{b}^{(2)}_2 ( \mu_2 ,  \mu_1)$ it follows that $\widetilde{\Psi} _2 ( \mu_1 ,  \mu_2)$ is symmetric with respect to the interchange of the variables $\mu_1$ and $\mu_2$.

Starting from the definitions \eqref{transfer-matrix} and \eqref{Psi2-til}, using the relations \eqref{ABBOmega}, \eqref{DBBOmega} and \eqref{CBBOmega}, from the appendix \ref{app: commut-rel}, to push the operators $\mathcal{A} (\lambda)$, $\widehat{\mathcal{D}} (\lambda)$ and $\mathcal{C} (\lambda)$ to the right and after rearranging some terms, we obtain the off-shell action of transfer matrix $t(\lambda)$ on $\widetilde{\Psi} _2 ( \mu_1 ,  \mu_2)$
\begin{align}
\label{t-on-Psi2-til}
t(\lambda) \widetilde{\Psi} _2 ( \mu_1 ,  \mu_2) &= \Lambda _2 (\lambda , \{\mu _i \} ) \widetilde{\Psi} _2 ( \mu_1 ,  \mu_2) 
+  \sum _{i=1}^2 \frac{ \sinh (\eta) \sinh (2 (\lambda + \eta ))}{\sinh (\lambda - \mu _i) \sinh (\lambda + \mu _i + \eta )} \times
\notag \\
&\times \kappa ^+ \sinh (\xi^+ - \mu _i ) \ F_2(\mu _i; \mu _{3-i}) \widetilde{\Psi} _2 ( \lambda ,  \mu_{3-i}) ,
\end{align}
where the eigenvalue is given by
\begin{align}
\label{Lambda2} 
\Lambda _2 (\lambda , \{\mu _i \}) &= \kappa _1 (\lambda) \ \alpha (\lambda) \ \prod _{i =1}^2
\frac{\sinh (\lambda + \mu_i) \sinh(\lambda - \mu_i - \eta)}{\sinh (\lambda - \mu _i) \sinh(\lambda + \mu_i + \eta)}  \notag \\[1ex]
&+ \kappa _2 (\lambda) \ \widehat{ \delta} (\lambda) \ \prod _{i =1}^2
\frac{\sinh (\lambda - \mu_i + \eta) \sinh (\lambda + \mu_i + 2 \eta )}{\sinh (\lambda - \mu_i ) \sinh (\lambda + \mu _i + \eta)} 
\end{align}
and the two unwanted terms in \eqref{t-on-Psi2-til} are canceled by the Bethe equations which follow from 
\begin{align}
\label{BE-2}
F_2(\mu _i; \mu _{3-i}) &= \frac{ \sinh (2\mu _i)}{\sinh (2 \mu _i  + \eta)} \
\frac{\sinh (\mu _ i + \mu _{3-i}) \sinh(\mu _ i - \mu _{3-i} - \eta )}{\sinh (\mu _ i - \mu _{3-i}) \sinh (\mu _ i + \mu _{3-i} + \eta )} \ \alpha (\mu _i) 
- \frac{\sinh (\xi^+ + \mu _i + \eta)}{\sinh (\xi^+ - \mu _i )} \times \notag \\
& \times \frac{\sinh (\mu _i - \mu_{3-i} + \eta) \sinh (\mu _i + \mu_{3-i} + 2 \eta)}{\sinh (\mu _i - \mu _{3-i}) \sinh (\mu _i + \mu_{3-i} + \eta)} \ \widehat{ \delta}  (\mu _i) = 0 ,  
\end{align}
with $i = \{ 1, 2 \}$. 
Therefore the Bethe equations are
\begin{equation}
\label{Bethe Eq-2.i}
\frac{\alpha (\mu _i)}{\widehat{ \delta} (\mu _i)} = \frac{\sinh (2 (\mu _i   + \eta)) \ \kappa _2 (\mu _i )}{\sinh (2 \mu _i ) \ \kappa _1 (\mu _i )} \
\frac{\sinh (\mu _i - \mu_{3-i} + \eta) \sinh (\mu _i + \mu_{3-i} + 2 \eta)}{\sinh (\mu _ i + \mu _{3-i}) \sinh (\mu _ i - \mu _{3-i} - \eta )} ,
\end{equation}
where $i = \{ 1, 2 \}$. This shows that $\widetilde{\Psi} _2 ( \mu_1 ,  \mu_2)$ \eqref{Psi2-til} is a Bethe vector of the transfer matrix \eqref{transfer-matrix} and, again, it is the one which in the scaling limit corresponds to the Bethe vector of the XXX chain \cite{CAMS}.

Furthermore, due to the following identities 
\begin{align}
\label{identity-Psi2-L1}
&\Lambda _1 (\lambda , \mu _2)  - \Lambda _2 (\lambda , \mu _1 , \mu _2) = \kappa ^+ \sinh ( \xi^+ - \lambda ) \frac{\sinh (\eta) \sinh ( 2 (\lambda + \eta) )}{\sinh (\lambda - \mu _1) \sinh(\lambda + \mu _1 + \eta)} F_2 (\lambda ; \mu _2) , \\[2ex]
\label{identity-Psi2-L2}
&\Lambda _1 (\lambda , \mu _1)  - \Lambda _2 (\lambda , \mu _1 , \mu _2) = \kappa ^+ \sinh ( \xi^+ - \lambda ) \frac{\sinh (\eta) \sinh ( 2 (\lambda + \eta) )}{\sinh (\lambda - \mu _2) \sinh(\lambda + \mu _2 + \eta)} F_2 (\lambda ; \mu _1) , \\[2ex]
\label{identity-Psi2-F1,2}
&\frac{F_2 (\mu _2 ; \mu _1) \ F_1  ( \mu _1) - F_2 (\mu _1 ; \mu _2) \ F_2 (\mu _2 ; \lambda)}{\sinh (\lambda - \mu _1) \sinh(\lambda + \mu _1 + \eta)} 
+ \frac{F_2 (\mu _1 ; \mu _2) \ F_1  ( \mu _2) - F_2 (\mu _2 ; \mu _1) \ F_2 (\mu _1 ; \lambda)}{\sinh (\lambda - \mu _2) \sinh(\lambda + \mu _2 + \eta)} = 0 ,
\end{align}  
the Bethe vector $\widetilde{\Psi }_2 (\mu_1 , \mu_2)$ \eqref{Psi2-til} can be generalized 
\begin{equation}
\label{Psi2-tilde}
\begin{split}
\widetilde{\Psi} _2 (\mu_1 , \mu_2, C_1, C_2) &= \widetilde{\Psi }_2 (\mu_1 , \mu_2) + C_2 \frac{\psi^+}{\kappa ^+}  \left( \sinh (\xi ^+ - \mu _1) F_2 (\mu _1 ; \mu _2)  \widetilde{\Psi }_1( \mu_2, C_1) \right. \\
& \left.  +  \sinh (\xi ^+ - \mu _2) F_2 (\mu _2 ; \mu _1) \widetilde{\Psi} _1( \mu_1, C_1) \right) ,
\end{split}
\end{equation}
where $C_2$ is independent of $\mu_1$ and  $\mu_2$ and $\widetilde{\Psi} _1( \mu_i, C_1)$ is the Bethe vector given in \eqref{Psi1-tilde}, so that the off-shell action of transfer matrix $t(\lambda)$ on $\widetilde{\Psi} _2 (\mu_1 , \mu_2, C_1, C_2)$ reads
\begin{align}
\label{t-on-Psi2-tilde}
t(\lambda) \widetilde{\Psi} _2 (\mu_1 , \mu_2, C_1, C_2) &= \Lambda _2 (\lambda , \{\mu _i \} ) \widetilde{\Psi} _2 (\mu_1 , \mu_2, C_1, C_2)
+  \sum _{i=1}^2 \frac{ \sinh (\eta) \sinh (2 (\lambda + \eta ))}{\sinh (\lambda - \mu _i) \sinh (\lambda + \mu _i + \eta )} \times
\notag \\
&\times \kappa ^+ \sinh (\xi^+ - \mu _i ) \ F_2(\mu _i; \mu _{3-i}) \widetilde{\Psi} _2 ( \lambda ,  \mu_{3-i}, C_1, C_2) .
\end{align}
Once more in \eqref{Psi2-tilde} we find that the general form of Bethe vectors can be expressed as a sum of a particular vector and a linear combination of lower order Bethe vectors. Due to  identities \eqref{identity-Psi2-L1}-\eqref{identity-Psi2-F1,2} this linear combination of lower order Bethe vectors corresponds the same eigenvalue as the particular vector \eqref{t-on-Psi2-tilde}. This is indeed the case with Bethe vectors of any order, for details see appendix \ref{app: B-vec}. To our knowledge, the existence of this freedom in the choice of the Bethe vector has hitherto remained unnoticed in the literature. In certain cases, it seems that omission to note this freedom can be traced to imposing, by some authors \cite{Lima13}, too strong requirements on the vanishing of the off-shell terms. Namely, all the terms (including vacuum ones) should be required to vanish only once the Bethe equations are imposed, and not necessarily to be identically zero.

However, in order to have the recurrence procedure for defining the higher order Bethe vectors it is instructive to set $C _1 = - \tanh (\eta), C _2 = 1$ in \eqref{Psi2-tilde} and to consider a particular Bethe vector 
\begin{equation}
\label{Psi2}
\begin{split}
\Psi _2 ( \mu_1 ,  \mu_2) &= \widetilde{\Psi} _2 (\mu_1 , \mu_2, C_1= - \tanh (\eta), C _2 = 1)  \\
&= \mathcal{B} (\mu_1) \mathcal{B} (\mu_2) \Omega _+ +  b^{(1)}_2(\mu_2 ; \mu_1)  \mathcal{B} (\mu_1) \Omega _+  +  b^{(1)}_2(\mu_1; \mu_2 ) \mathcal{B} (\mu_2) \Omega _+ + b^{(2)}_2 ( \mu_1 ,  \mu_2) \Omega _+ ,
\end{split}
\end{equation}
where the functions $b^{(1)}_2(\mu_1; \mu_2 )$ and $b^{(2)}_2( \mu_1 ,  \mu_2)$ are given by
\begin{align}
\label{b2-1}
b^{(1)}_2 (\mu_1; \mu_2 ) &= \left( - \frac{\psi^+}{\kappa ^+} \right) \left( \frac{\sinh (\mu _1 + \mu _2) \sinh (\mu _1 - \mu _2 - \eta )}{\sinh (\mu _1 - \mu _2) \sinh (\mu _1 + \mu _2 + \eta )}   \frac{ \sinh (2\mu _1)}{\sinh (2 \mu _1  + \eta)}  e^{-(\xi ^+ - \mu _1)} \ \alpha (\mu _1 ) \right.
\notag \\[1ex]
& \left. \  -  \frac{\sinh (\mu _1 - \mu_2 + \eta) \sinh (\mu _1 + \mu_2 + 2 \eta)}{\sinh (\mu _1 - \mu _2) \sinh (\mu _1 + \mu_2 + \eta)} \ e^{-( \xi ^+ + \mu _1 + \eta)} \ \widehat{ \delta}  (\mu _1) \right ) ,
\end{align}
and
\begin{align}
\label{b2-2}
b^{(2)}_2 ( \mu_1 ,  \mu_2) &=\left( \frac{\psi^+}{\kappa ^+} \right) ^2 \left( \frac{\sinh (\mu _1 + \mu_2)}{\sinh (\mu _1 + \mu _2 + \eta)} 
\frac{\sinh (2 \mu _1)}{\sinh (2 \mu _1 + \eta)} \, \frac{\sinh (2 \mu _2)}{\sinh (2 \mu _2 + \eta)} \ e^{ -(2 \xi ^+ - \mu _1 - \mu _2 + \eta )} \alpha (\mu _1) \alpha (\mu _2) \right.
\notag \\[1ex]
%
%
&- \frac{\sinh (2 \mu _1)}{\sinh (2 \mu _1 + \eta)} \, \frac{\sinh (\mu_1 - \mu_2 - \eta)}{\sinh (\mu_1 - \mu_2)}  \ e^{ - (2 \xi ^+ - \mu _1 + \mu _2 + 2 \eta )} \ \alpha (\mu _1)   \widehat{ \delta}  (\mu _2) \notag \\[1ex]
&- \frac{\sinh (\mu_2 - \mu_1 - \eta)}{\sinh (\mu_2 - \mu_1)}  \, \frac{\sinh (2 \mu _2)}{\sinh (2 \mu _2 + \eta)} \ e^{ - ( 2 \xi ^+ + \mu _1 - \mu _2 + 2 \eta )} \ \widehat{ \delta} (\mu _1) \alpha (\mu _2) \notag \\[1ex]
&\left. + \frac{\sinh (\mu_1 + \mu_2 + 2 \eta)}{\sinh (\mu_1 + \mu_2 + \eta)} \ e^{ - (2 \xi ^+ + \mu _1 + \mu _2 + 3 \eta ) } \ \widehat{ \delta} (\mu _1) \widehat{ \delta} (\mu _2) \right) .
\end{align}
A key observation here is that the above function  $b^{(2)}_2 ( \mu_1 ,  \mu_2)$ can be expressed in terms of the coefficient functions $b^{(1)}_2 (\mu_1; \mu_2 )$ \eqref{b2-1} and $b_ 1 (\mu_i)$ \eqref{b1} as follows
\begin{equation}
\label{b2-2-rec}
b^{(2)}_2 ( \mu_1 ,  \mu_2) = \frac{1}{1 + e^{2\eta}} \left(  b^{(1)}_2 (\mu_1; \mu_2 ) \, b_1 (\mu _2) +
b^{(1)}_2 (\mu_2; \mu_1 )  \, b_1 (\mu _1) \right) .
\end{equation} 
This relation is essential in the recurrence procedure for obtaining general form of the Bethe vectors. It coincides, up to the multiplicative factor, with the recurrence relation defining the function $b^{(2)}_2 ( \mu_1 ,  \mu_2)$ in the case of the corresponding Bethe vector of the XXX Heisenberg spin chain, the equation (V.25) in \cite{CAMS}. 

Although, as we have seen, the Bethe vectors $\Psi_1 (\mu)$ \eqref{Psi1} and $\Psi _2 ( \mu_1 ,  \mu_2)$ \eqref{Psi2} correspond to the particular choice of parameters $C_i$ in \eqref{Psi1-tilde} and \eqref{Psi2-tilde}, respectively,  it turns out that these vectors admit the recurrence procedure analogous to the one applied in the case of the XXX Heisenberg spin chain  \cite{CAMS}. Before addressing the general case of the Bethe vector $\Psi _M ( \mu_1 ,  \mu_2 ,  \dots ,  \mu_M )$, for an arbitrary positive integer $M$, we will present below the $M=3$ case as an insightful example. The Bethe vector $\Psi _3 ( \mu_1 ,  \mu_2 ,  \mu_3)$ we propose is a symmetric function of its arguments and it is given as the following sum of eight terms
\begin{equation} 
\label{Psi3}
\begin{split}
\Psi _3 ( \mu_1 ,  \mu_2 ,  \mu_3) = \mathcal{B} (\mu_1) \mathcal{B} (\mu_2) \mathcal{B} (\mu_3)\Omega _+ 
+  b ^{(1)}_3(\mu_3 ; \mu_2 , \mu_1) \mathcal{B} (\mu_1) \mathcal{B} (\mu_2) \Omega _+ 
+ b ^{(1)}_3(\mu_1 ;  \mu_2 ,  \mu_3)  \times \\
\times \mathcal{B} (\mu_2) \mathcal{B} (\mu_3) \Omega _+ 
+ b ^{(1)}_3(\mu_2 ; \mu_1 , \mu_3) \mathcal{B} (\mu_1) \mathcal{B} (\mu_3) \Omega _+ 
+ b ^{(2)}_3(\mu_1 ,  \mu_2 ;  \mu_3)  \mathcal{B} (\mu_3) \Omega _+  \\
+ b ^{(2)}_3(\mu_1, \mu_3 ;  \mu_2) \mathcal{B} (\mu_2) \Omega _+ 
+ b ^{(2)}_3(\mu_2, \mu_3 ;  \mu_1) \mathcal{B} (\mu_1) \Omega _+ 
+ b ^{(3)}_3 ( \mu_1 ,  \mu_2 ,  \mu_3) \Omega _+ ,
\end{split}
\end{equation}
where the coefficient functions $b ^{(1)}_3 (\mu_1 ;  \mu_2 ,  \mu_3)$ , $b ^{(2)}_3 (\mu_1 ,  \mu_2 ;  \mu_3)$ and $b ^{(3)}_3 (\mu_1 ,  \mu_2 ,  \mu_3)$ are given by
\begin{align}
\label{b3-1}
b ^{(1)}_3 (\mu_1 ;  \mu_2 ,  \mu_3) &= \left( - \frac{\psi^+}{\kappa ^+} \right)
 \left( \prod _{i=2}^3 \frac{\sinh (\mu _1 + \mu _i) \sinh (\mu _1 - \mu _i - \eta )}{\sinh (\mu _1 - \mu _i) \sinh (\mu _1 + \mu _i + \eta )} \ \frac{ \sinh (2 \mu _1)}{\sinh (2 \mu _1  + \eta)}
\ e^{-(\xi ^+ - \mu _1)} \ \alpha (\mu _1) \right. \notag \\[1ex]
&\left. - \prod _{i=2}^3 \frac{\sinh (\mu _1 - \mu_i + \eta) \sinh (\mu _1 + \mu_i + 2 \eta)}{\sinh (\mu _1 - \mu _i) \sinh (\mu _1 + \mu_i + \eta)} \ e^{- (\xi ^+ + \mu _1 + \eta)}  \ \widehat{ \delta}  (\mu _1) \right)  ,  
\end{align}
\begin{align}
\label{b3-2}
b ^{(2)}_3 (\mu_1 ,  \mu_2 ;  \mu_3) &= \frac{1}{1 + e^{2\eta}} \left( b^{(1)}_3 (\mu_1; \mu_2 , \mu_3 )  \, b^{(1)}_2 (\mu_2; \mu_3 ) +b^{(1)}_3 (\mu_2; \mu_1 , \mu_3 )  \, b^{(1)}_2 (\mu_1; \mu_3 ) \right) ,  \\[1ex]
\label{b3-3}
b ^{(3)}_3 (\mu_1 ,  \mu_2 ,  \mu_3) &=  \frac{1}{1+2e^{2\eta}+2e^{4\eta}+e^{6\eta}} \left(  b^{(1)}_3 (\mu_1; \mu_2 , \mu_3 )  \, b^{(1)}_2 (\mu_2; \mu_3 ) \, b_1 (\mu _3) \right. \notag \\[1ex]
&+  b^{(1)}_3 (\mu_1; \mu_2 , \mu_3 )  \, b^{(1)}_2 (\mu_3 ; \mu_2 ) \, b_1 (\mu _2) 
+ b^{(1)}_3 (\mu_2; \mu_1 , \mu_3 )  \, b^{(1)}_2 (\mu_1 ; \mu_3 ) \, b_1 (\mu _3)
\notag \\[1ex]
&+ b^{(1)}_3 (\mu_2; \mu_1 , \mu_3 )  \, b^{(1)}_2 (\mu_3 ; \mu_1 ) \, b_1 (\mu _1) 
+ b^{(1)}_3 (\mu_3 ; \mu_1 , \mu_2 )  \, b^{(1)}_2 (\mu_1 ; \mu_2 ) \, b_1 (\mu _2)
\notag \\[1ex]
&\left. + b^{(1)}_3 (\mu_3; \mu_1 , \mu_2 )  \, b^{(1)}_2 (\mu_2 ; \mu_1 ) \, b_1 (\mu _1) \right) .
\end{align}
It is important to notice that the coefficient functions $b ^{(2)}_3 (\mu_1 ,  \mu_2 ;  \mu_3)$ and $b ^{(3)}_3 (\mu_1 ,  \mu_2 ,  \mu_3)$ are defined above in terms of the function $b ^{(1)}_3 (\mu_1 ;  \mu_2 ,  \mu_3)$ and the functions $b^{(1)}_2 (\mu_1; \mu_2)$ and $b_1 (\mu)$ already given in \eqref{b2-1} and \eqref{b1}, respectively. The action of $t(\lambda)$ \eqref{transfer-matrix} on $\Psi _3 ( \mu_1 ,  \mu_2 ,  \mu_3)$, obtained using evident generalization of the formulas \eqref{ABBOmega}, \eqref{DBBOmega} and  \eqref{CBBOmega} and subsequent rearranging of terms, reads
\begin{equation}
\label{t-on-Psi3}
\begin{split} 
t(\lambda) \Psi _3 ( \mu_1 ,  \mu_2 ,  \mu_3) &= \Lambda _3 (\lambda , \{\mu _i \}) \Psi _3 ( \mu_1 ,  \mu_2 ,  \mu_3) + \sum _{i=1}^3 \frac{ \sinh (\eta) \sinh (2 (\lambda + \eta ))}{\sinh (\lambda - \mu _i) \sinh (\lambda + \mu _i + \eta )} \times \\[1ex]
&\times \ \kappa ^+ \sinh (\xi^+ - \mu _i ) \ F_3 (\mu _i ; \{\mu _j \} _{j \neq i}) \ \Psi _3 ( \lambda ,  \{\mu _j \} _{j \neq i}) ,
\end{split}
\end{equation}
where the eigenvalue is given by
\begin{align}
\label{Lambda3} 
\Lambda _3 (\lambda , \{\mu _i \}) &= \kappa _1 (\lambda) \ \alpha (\lambda) \ \prod _{i =1}^3
\frac{\sinh (\lambda + \mu_i) \sinh(\lambda - \mu_i - \eta)}{\sinh (\lambda - \mu _i) \sinh(\lambda + \mu_i + \eta)}  \notag \\[1ex]
&+ \kappa _2 (\lambda) \ \widehat{ \delta} (\lambda) \ \prod _{i =1}^3
\frac{\sinh (\lambda - \mu_i + \eta) \sinh (\lambda + \mu_i + 2 \eta )}{\sinh (\lambda - \mu_i ) \sinh (\lambda + \mu _i + \eta)} 
\end{align}
and the three unwanted terms in \eqref{t-on-Psi3} are canceled by the Bethe equations which follow from 
\begin{align}
\label{BE-3}
F_3 (\mu _i ; \{\mu _j \} _{j \neq i}) &= \frac{ \sinh (2\mu _i)}{\sinh (2 \mu _i  + \eta)} \ \alpha (\mu _i) \ \prod _{\substack{ j =1\\ j \neq i}}^3 
\frac{\sinh (\mu _ i + \mu _j) \sinh(\mu _ i - \mu _j - \eta )}{\sinh (\mu _ i - \mu _j ) \sinh (\mu _ i + \mu _j + \eta )}  \notag \\[1ex]
& - \frac{\sinh (\xi^+ + \mu _i + \eta)}{\sinh (\xi^+ - \mu _i )} \ \widehat{ \delta}  (\mu _i) \ \prod _{\substack{ j =1\\ j \neq i}}^3 \frac{\sinh (\mu _i - \mu_j + \eta) \sinh (\mu _i + \mu_j + 2 \eta)}{\sinh (\mu _i - \mu _j ) \sinh (\mu _i + \mu_j + \eta)}  = 0 ,  
\end{align}
with $i = \{ 1, 2 , 3 \}$. 
Therefore the Bethe equations are
\begin{equation}
\label{Bethe Eq-3.i}
\frac{\alpha (\mu _i)}{\widehat{ \delta} (\mu _i)} = \frac{\sinh (2 (\mu _i   + \eta)) \ \kappa _2 (\mu _i )}{\sinh (2 \mu _i ) \ \kappa _1 (\mu _i )} \
\prod _{\substack{ j =1\\ j \neq i}}^3  \frac{\sinh (\mu _i - \mu_j + \eta) \sinh (\mu _i + \mu_j + 2 \eta)}{\sinh (\mu _ i + \mu _j) \sinh (\mu _ i - \mu _j - \eta )} ,
\end{equation}
where $i = \{ 1, 2 , 3 \}$. Thus, as expected, we have obtained the strikingly simple expression for the off-shell action of the transfer matrix of the XXZ Heisenberg chain with the upper triangular reflection matrices on the Bethe vector $\Psi _3 ( \mu_1 ,  \mu_2 ,  \mu_3 )$, which is by definition \eqref{Psi3} symmetric function of its arguments $\{\mu _i \}_{I=1}^3$. As before, $\Psi _3 ( \mu_1 ,  \mu_2 ,  \mu_3 )$ is a special case of the more general Bethe vector $\widetilde{\Psi }_3 ( \mu_1 ,  \mu_2, \mu_3, C_1, C_2, C_3)$ we have found along the lines similar to the  $M=1$ and $M=2$ cases, for details see the appendix \ref{app: B-vec}, where we also give the generalized form of the Bethe vector for arbitrary M. The most significant advantage of this particular form of the Bethe vector is that it is defined by the recurrence procedure which is analogous to the one proposed in the case of the XXX Heisenberg  chain \cite{CAMS}. Notice the right-hand-side of the equations \eqref{b3-2} and \eqref{b3-3} differ only by the multiplicative factors from the analogous equations (V.32) and (V.34) in \cite{CAMS}.

We readily proceed to define $\Psi _M ( \mu_1 ,  \mu_2 ,  \dots ,  \mu_M )$ as a sum of $2^M$ terms, for an arbitrary positive integer $M$, and as a symmetric function of its arguments by the recurrence procedure
\begin{equation}
\label{PsiM}
\begin{split}
&\Psi _M ( \mu_1 ,  \mu_2 ,  \dots ,  \mu_M ) = \mathcal{B} (\mu_1) \mathcal{B} (\mu_2) \cdots \mathcal{B} (\mu_M) \Omega _+ 
+  b ^{(1)}_M(\mu_M ; \mu_1 ,  \mu_2 ,  \dots ,  \mu_{M-1})\mathcal{B} (\mu_1) \mathcal{B} (\mu_2) \cdots \mathcal{B} (\mu_{M-1}) \Omega _+  \\
&+\cdots + b ^{(2)}_M(\mu_{M-1} , \mu_M  ; \mu_1 ,  \mu_2 ,  \dots ,  \mu_{M-2}) \mathcal{B} (\mu_1) \mathcal{B} (\mu_2) \cdots \mathcal{B} (\mu_{M-2}) \Omega _+   \\
&\ \ \vdots   \\
&+ b ^{(M-1)}_M (\mu_1 ,  \mu_2 ,  \dots ,  \mu_{M-1} ;  \mu_M) \mathcal{B} (\mu_M)  \Omega _+
+ b ^{(M)}_M ( \mu_1 ,  \mu_2 ,  \dots ,  \mu_M) \Omega _+ ,
\end{split}
\end{equation}
where the first coefficient function is explicitly given  by
\begin{equation}
\label{bM-1}
\begin{split}
 &b ^{(1)}_M(\mu_1 ;  \mu_2 ,  \mu_3 ,  \dots ,  \mu_M) = \left( - \frac{\psi^+}{\kappa ^+} \right)
 \left( \prod _{i=2}^M \frac{\sinh (\mu _1 + \mu _i) \sinh (\mu _1 - \mu _i - \eta )}{\sinh (\mu _1 - \mu _i) \sinh (\mu _1 + \mu _i + \eta )} \ \frac{ \sinh (2 \mu _1)}{\sinh (2 \mu _1  + \eta)}
\ \times \right. \\[1ex]
&\left. \times \ e^{-(\xi ^+ - \mu _1)} \ \alpha (\mu _1) - \prod _{i=2}^M \frac{\sinh (\mu _1 - \mu_i + \eta) \sinh (\mu _1 + \mu_i + 2 \eta)}{\sinh (\mu _1 - \mu _i) \sinh (\mu _1 + \mu_i + \eta)} \ e^{- (\xi ^+ + \mu _1 + \eta)}  \ \widehat{ \delta}  (\mu _1) \right)  , 
\end{split}
\end{equation}
and all the other coefficient functions are given by the following recurrence formulae
\begin{align}
\label{bM-2}
b ^{(2)}_M(\mu_1 , \mu_2 ;  \mu_3 ,  \dots ,  \mu_M) &= \frac{q^{-1}}{[2]_q !} \left(  b ^{(1)}_M(\mu_1 ;  \mu_2 ,  \mu_3 ,  \dots ,  \mu_M) b ^{(1)}_{M-1}(\mu_2 ;  \mu_3 ,  \dots ,  \mu_M) \right. \notag \\[1ex]
&\left. +  b ^{(1)}_M(\mu_2 ;  \mu_1 ,  \mu_3 ,  \dots ,  \mu_M) b ^{(1)}_{M-1}(\mu_1 ;  \mu_3 ,  \dots ,  \mu_M)
\right) , \\
&\ \ \vdots \notag \\
\label{bM-(M-1)}
b ^{(M-1)}_M (\mu_1 ,  \mu_2 ,  \dots ,  \mu_{M-1} ;  \mu_M) &=  \frac{q^{-\frac{(M-1)(M-2)}{2}}}{[M-1]_q !} \sum_{\rho \in S_{M-1} } b^{(1)}_M (\mu_{\rho (1)}; \mu_{\rho (2)} , \dots , \mu_{M} )  \, b ^{(1)}_{M-1}(\mu_{\rho (2)} ;  \mu_{\rho (3)} ,  \dots ,  \mu_M) \times \notag \\[1ex]
&\times b ^{(1)}_{M-2}(\mu_{\rho (3)} ;  \mu_{\rho (4)} ,  \dots ,  \mu_M) \cdots b^{(1)}_2 ( \mu_{\rho (M-1)} ; \mu_{M}) \\[2ex]
\label{bM-M}
b ^{(M)}_M ( \mu_1 ,  \mu_2 ,  \dots ,  \mu_M) &=   \frac{q^{-\frac{M(M-1)}{2}}}{[M]_q !} \sum_{\sigma \in S_M} b^{(1)}_M (\mu_{\sigma (1)}; \mu_{\sigma (2)} , \dots , \mu_{\sigma (M)} )  \, b ^{(1)}_{M-1}(\mu_{\sigma (2)} ;  \mu_{\sigma (3)} ,  \dots ,  \mu_{\sigma (M)}) \times \notag \\[1ex]
&\times b ^{(1)}_{M-2}(\mu_{\sigma (3)} ;  \mu_{\sigma (4)} ,  \dots , \mu_{\sigma (M)}) \cdots b^{(1)}_2 ( \mu_{\sigma (M-1)} ;\mu_{\sigma (M)}) \, b_1 (\mu_{\sigma (M)}) ,
\end{align}
where, for a positive integer  $N$, $[N]_q = \frac{q^N - q^{-N}}{q - q^{-1}}$ and $[N]_q ! = [N]_q \cdot [N-1]_q \cdots [2]_q \cdot [1]_q $, with $q = e^{\eta}$ and $S_{M-1}$ and $S_M$ are the symmetric groups of degree $M-1$ and $M$, respectively. As is the case $M=3$, the formulae \eqref{bM-2}-\eqref{bM-M} are deformation of the corresponding relations (V.32) - (V.35) in the case of the XXX Heisenberg  chain \cite{CAMS}.

A straightforward calculation based on evident generalization of the formulas \eqref{ABBOmega}, \eqref{DBBOmega} and  \eqref{CBBOmega} and subsequent rearranging of terms, yields
the off-shell action of the transfer matrix on the Bethe vector $\Psi _M ( \mu_1 ,  \mu_2 ,  \dots ,  \mu_M )$ 
\begin{equation}
\label{t-on-PsiM}
\begin{split} 
t(\lambda) \Psi _M ( \mu_1 ,  \mu_2 ,  \dots ,  \mu_M ) &= \Lambda _M (\lambda , \{\mu _i \}) \Psi _M ( \mu_1 ,  \mu_2 ,  \dots ,  \mu_M )  + \sum _{i=1}^M \frac{ \sinh (\eta) \sinh (2 (\lambda + \eta ))}{\sinh (\lambda - \mu _i) \sinh (\lambda + \mu _i + \eta )} \times \\[1ex]
&\times \ \kappa ^+ \sinh (\xi^+ - \mu _i ) \ F_M (\mu _i ; \{\mu _j \} _{j \neq i}) \ \Psi _M ( \lambda ,  \{\mu _j \} _{j \neq i}) ,
\end{split}
\end{equation}
where the corresponding eigenvalue is given by
\begin{align}
\label{LambdaM} 
\Lambda _M (\lambda , \{\mu _i \}) &= \kappa _1 (\lambda) \ \alpha (\lambda) \ \prod _{i =1}^M
\frac{\sinh (\lambda + \mu_i) \sinh(\lambda - \mu_i - \eta)}{\sinh (\lambda - \mu _i) \sinh(\lambda + \mu_i + \eta)}  \notag \\[1ex]
&+ \kappa _2 (\lambda) \ \widehat{ \delta} (\lambda) \ \prod _{i =1}^M
\frac{\sinh (\lambda - \mu_i + \eta) \sinh (\lambda + \mu_i + 2 \eta )}{\sinh (\lambda - \mu_i ) \sinh (\lambda + \mu _i + \eta)} 
\end{align}
and the $M$ unwanted terms on the right hand side of \eqref{t-on-PsiM} are canceled by the Bethe equations which follow from 
\begin{align}
\label{Be-M}
F_M (\mu _i ; \{\mu _j \} _{j \neq i}) &= \frac{ \sinh (2\mu _i)}{\sinh (2 \mu _i  + \eta)} \ \alpha (\mu _i) \ \prod _{\substack{ j =1\\ j \neq i}}^M 
\frac{\sinh (\mu _ i + \mu _j) \sinh(\mu _ i - \mu _j - \eta )}{\sinh (\mu _ i - \mu _j ) \sinh (\mu _ i + \mu _j + \eta )}  \notag \\[1ex]
& - \frac{\sinh (\xi^+ + \mu _i + \eta)}{\sinh (\xi^+ - \mu _i )} \ \widehat{ \delta}  (\mu _i) \ \prod _{\substack{ j =1\\ j \neq i}}^M \frac{\sinh (\mu _i - \mu_j + \eta) \sinh (\mu _i + \mu_j + 2 \eta)}{\sinh (\mu _i - \mu _j ) \sinh (\mu _i + \mu_j + \eta)}  = 0 ,  
\end{align}
with $i = \{ 1, 2 , \ldots , M \}$. Therefore the Bethe equations are
\begin{equation}
\label{Bethe Eq-3.i}
\frac{\alpha (\mu _i)}{\widehat{ \delta} (\mu _i)} = \frac{\sinh (2 (\mu _i   + \eta)) \ \kappa _2 (\mu _i )}{\sinh (2 \mu _i ) \ \kappa _1 (\mu _i )} \
\prod _{\substack{ j =1\\ j \neq i}}^M  \frac{\sinh (\mu _i - \mu_j + \eta) \sinh (\mu _i + \mu_j + 2 \eta)}{\sinh (\mu _ i + \mu _j) \sinh (\mu _ i - \mu _j - \eta )} ,
\end{equation}
where $i = \{ 1, 2 , \ldots , M \}$.  The Bethe vector $\Psi _M ( \mu_1 ,  \mu_2 ,  \dots ,  \mu_M )$ we have defined in \eqref{PsiM} yields the strikingly simple expression \eqref{t-on-PsiM} for the off-shell action of the transfer matrix $t(\lambda)$ \eqref{transfer-matrix}. Thus we have fully implemented the algebraic Bethe ansatz for the XXZ Heisenberg spin chain with the triangular reflection matrices. In the following section, we will explored further these results through the so-called quasi-classical limit in order to investigate the corresponding Gaudin model \cite{CAMRS}.

\section{Corresponding Gaudin model \label{sec: Gm}}
As it is well known \cite{CAMS,CAMRS,CAMN,MS2}, the study of the open Gaudin model requires that the parameters of the reflection matrices on the left and on the right end of the chain are the same. Thus, we impose
\begin{equation}
\label{normalizationKpl}
\lim_{\eta \to 0}\Big(  K^+(\lambda) K^{-} (\lambda)\Big)  = \kappa ^2 \sinh (\xi - \lambda) \sinh (\xi + \lambda)  \mathbbm{1}.
\end{equation}
Notice that in general this not the case in  the study of the open spin chain. However, this condition is essential for the Gaudin model. Therefore we will write
\begin{equation}
\label{K-min-Gm}
K^-(\lambda)\equiv K(\lambda) = \left(\begin{array}{cc}
\kappa  \sinh (\xi  + \lambda) & \psi  \sinh (2 \lambda) \\ 
0 & \kappa \sinh (\xi  - \lambda) \end{array}\right) 
\end{equation}
so that 
\begin{equation} 
\label{K-plus-Gm}
K^+(\lambda)= K(-\lambda-\eta)= \left(\begin{array}{cc}
\kappa \sinh ( \xi - \lambda - \eta) & - \psi  \sinh \left( 2 (\lambda + \eta) \right) \\ 
0 & \kappa \sinh (\xi  + \lambda + \eta)
\end{array}\right) .
\end{equation}

In \cite{MS2} we have derived the generating function of the trigonometric Gaudin Hamiltonians with boundary terms following the approach of Sklyanin in the periodic case \cite{Sklyanin89, MNS}. Analogously to the rational case \cite{CAMRS, CAMS}, our derivation is based on the quasi-classical expansion of the linear combination of the transfer matrix of the XXZ chain and the central element, the so-called Sklyanin determinant. Finally, the expansion reads \cite{MS2} 
\begin{align}
\label{exp-difference}
t (\lambda) - \frac{\Delta \left[\mathcal{T}(\lambda) \right] }{\sinh (2\lambda)}  &= \kappa ^2 \sinh (\xi + \lambda ) \sinh (\xi - \lambda ) \mathbbm{1}
+ \eta \ \frac{\kappa ^2}{2}  \left( \cosh (2 \xi)  \coth (2 \lambda) - \frac{\cosh(4 \lambda)}{\sinh(2 \lambda)} \right)  \mathbbm{1} 
\notag \\[1ex]
&+ \frac{\eta ^2}{2}  \ \kappa ^2  \left( \sinh (\xi + \lambda ) \sinh (\xi - \lambda ) \left( \tau (\lambda) + \mathbbm{1} \right) - \frac{1}{2} \cosh (2 \lambda ) \mathbbm{1} \right) + \mathcal{O}(\eta ^3) ,
\end{align}
where $\tau (\lambda)$ is the generating function of the trigonometric Gaudin Hamiltonians with boundary terms
\begin{equation}
\label{b-tau} 
\tau (\lambda) =  \mathrm{tr}_0 \, \mathcal{L}_0 ^2(\lambda) ,
\end{equation}
 where 
\begin{equation}
\label{cal-L}
\mathcal{L} _0 (\lambda) = L _0 (\lambda) - K_0 (\lambda)  L _0 (- \lambda) K_0^{-1}(\lambda), \\
\end{equation}
with the Gaudin Lax matrix defined by 
\begin{equation}
\label{L-Gaudin}
L_0(\lambda) = \sum _{m=1}^N  \left( \sigma_{0}^3 \otimes \coth (\lambda - \alpha _m) S_{m}^3 + \frac{\sigma_{0}^+ \otimes S_{m}^- + \sigma_{0}^- \otimes S_{m}^+}{2 \, \sinh (\lambda - \alpha _m)} \right) , 
\end{equation}
and $K_0 (\lambda)$ the upper triangular reflection matrix given in \eqref{K-min-Gm}. The trigonometric Gaudin Hamiltonians with the boundary terms are obtained from the residues of the generating function $\tau (\lambda)$ \eqref{b-tau} at  poles $\lambda = \pm\alpha_m$ :
\begin{equation}
\label{res-Ham}
\underset {\lambda = \alpha_m} {\mbox{Res}} \tau (\lambda) \ = \  4 \, H_m
\mb{and}
\underset {\lambda = -\alpha_m} {\mbox{Res}} \tau (\lambda) \ =\ (-4) \,  H_m
\end{equation}
where
\begin{align}
\label{open-Ham-a}
H_m &= \sum _{n \neq m}^N \left( \coth (\alpha _m - \alpha _n)  \ S^3_m S^3_n + \frac{S^+_m S^-_n + S^-_m S^+_n}{2 \sinh (\alpha _m - \alpha _n)} \right) + \sum _{n=1}^N \coth (\alpha _m + \alpha _n) \
\frac{S^3_m S^3_n + S^3_n S^3_m}{2} \notag \\[1ex]
&+ \frac{\psi}{\kappa} \ \frac{\sinh (2 \alpha _m)}{\sinh(\xi + \alpha_m)}  \sum _{n=1}^N \frac{S^3_m S^+_n + S^+_n S^3_m}{2 \sinh (\alpha _m + \alpha _n)} + \frac{\sinh (\xi - \alpha_m) }{2 \sinh (\xi + \alpha_m)} \sum _{n=1}^N \frac{S^-_m S^+_n + S^+_n S^-_m}{2 \sinh (\alpha _m + \alpha _n)}
\notag \\[1ex]
&- \frac{\psi}{\kappa} \ \frac{\sinh (2 \alpha _m)}{\sinh(\xi - \alpha_m)}  \sum _{n=1}^N \coth (\alpha _m + \alpha _n) \ \frac{S^+_m S^3_n + S^3_n S^+_m}{2} + \frac{\sinh (\xi + \alpha_m) }{2 \sinh (\xi - \alpha_m)} \sum _{n=1}^N \frac{S^+_m S^-_n + S^-_n S^+_m}{2 \sinh (\alpha _m + \alpha _n)}
\notag \\[1ex]
&- \frac{\psi ^2}{\kappa ^2} \ \frac{\sinh ^2 (2 \alpha _m )}{2 \sinh (\xi - \alpha _m) \sinh (\xi + \alpha _m)}
\sum _{n=1}^N \frac{S^+_m S^+_n + S^+_n S^+_m}{2 \sinh (\alpha _m + \alpha _n)} .
\end{align} 

Since the central element $\Delta \left[\mathcal{T}(\lambda)\right]$ can be expressed in form \eqref{Del-calT}
it is evident that the vector $\Omega _+$ \eqref{Omega+} is its eigenvector
\begin{equation}
\label{Del-calTOm}
\Delta \left[\mathcal{T}(\lambda)\right] \Omega _+ =  \sinh (2 \lambda) \ \alpha (\lambda + \eta / 2) \, \widehat{\delta} (\lambda - \eta / 2) \, \Omega _+ .
\end{equation}
Moreover, it follows from \eqref{t-on-Om+} and \eqref{Del-calTOm} that $\Omega _+$ \eqref{Omega+}  is an eigenvector of the difference
\begin{equation}
\label{t-DonOm}
\left( t (\lambda) - \frac{\Delta \left[\mathcal{T}(\lambda) \right] }{\sinh (2\lambda)}  \right) \Omega _+ = \left( \Lambda _0 (\lambda) - \alpha (\lambda + \eta / 2) \, \widehat{\delta} (\lambda - \eta / 2) \right) \Omega _+ .
\end{equation}
We can expand the eigenvalue on the right hand side of the equation above in powers of $\eta$, taking into account that $\phi = 0$, 
\begin{align}
\label{exp-chi0}
&\left(  \kappa _1 (\lambda) \alpha (\lambda) + \kappa _2 (\lambda) \widehat{\delta} (\lambda) - \alpha (\lambda + \eta / 2) \, \widehat{\delta} (\lambda - \eta / 2) \right) 
=  \kappa ^2 \sinh (\xi + \lambda ) \sinh (\xi - \lambda )  \notag \\
&+ \eta \ \frac{\kappa ^2}{2}  \left( \cosh (2 \xi)  \coth (2 \lambda) - \frac{\cosh(4 \lambda)}{\sinh(2 \lambda)} \right)  
\notag\\
&+ \frac{\eta ^2}{2}  \ \kappa ^2  \left( \sinh (\xi + \lambda ) \sinh (\xi - \lambda ) \left( \chi _0 (\lambda) + 1 \right) - \frac{1}{2} \cosh (2 \lambda ) \right) + \mathcal{O}(\eta ^3) .  
\end{align}
Substituting the expansion above into the right hand side of \eqref{t-DonOm} and using \eqref{exp-difference} to expand the left hand side, it follows that the vector $\Omega _+$ \eqref{Omega+} is an eigenvector of the generating function of the Gaudin Hamiltonians
\begin{equation}
\label{egnv-chi0}
\tau (\lambda) \Omega _+ = \chi _0 (\lambda)  \Omega _+,
\end{equation}
with
\begin{equation}
\label{chi0}
\begin{split}
\chi _0 (\lambda) &=  2 \sum _{m,n = 1}^N  \left( s _m s _n + \frac{\sinh (\xi + \alpha _m ) \sinh (\xi - \alpha _m )}{\sinh (\xi + \lambda ) \sinh (\xi - \lambda )} \ s_m \delta _{mn} \right) 
\times \\
& \times 
\left( \coth (\lambda - \alpha _m) \coth (\lambda - \alpha _n) + 2 \coth (\lambda - \alpha _m) 
\coth (\lambda + \alpha _n)  
+  \coth (\lambda + \alpha _m) \coth (\lambda + \alpha _n) \right) .
\end{split}
\end{equation}
Moreover we can obtain the spectrum of the generating function of the Gaudin Hamiltonians through the expansion
\begin{align}
\label{exp-LambdaM}
&\left( \Lambda _M (\lambda, \{\mu_j \}_{j=1}^M ) - \alpha (\lambda + \eta / 2) \, \widehat{\delta} (\lambda - \eta / 2) \right) 
=  \kappa ^2 \sinh (\xi + \lambda ) \sinh (\xi - \lambda )  \notag \\
&+ \eta \ \frac{\kappa ^2}{2} \left( \cosh (2 \xi)  \coth (2 \lambda) - \frac{\cosh(4 \lambda)}{\sinh(2 \lambda)} \right)  
\notag\\
&+ \frac{\eta ^2}{2}  \ \kappa ^2  \left( \sinh (\xi + \lambda ) \sinh (\xi - \lambda ) \left( \chi _M (\lambda, \{\mu_j \}_{j=1}^M ) + 1 \right) - \frac{1}{2} \cosh (2 \lambda ) \right) + \mathcal{O}(\eta ^3) ,  
\end{align}
where
\begin{equation}
\label{chiM}
\begin{split}
\chi _M ( \lambda, \{\mu_j \}_{j=1}^M ) &= \frac{-2 \sinh(2\lambda)}{\sinh(\xi - \lambda) \sinh(\xi + \lambda)} \ 
\sum _{j=1}^M \frac{\sinh(2\lambda)}{\sinh (\lambda - \mu_j) \sinh (\lambda + \mu_j)} \\[1ex]
&+ 4 \sum _{j=1}^{M-1} \sum _{k=j+1}^M \frac{ \sinh(2\lambda)}{\sinh (\lambda - \mu_j) \sinh (\lambda + \mu_j)} \frac{ \sinh(2\lambda)}{\sinh (\lambda - \mu_k) \sinh (\lambda + \mu_k)} \\[1ex]
&- 4 \sum _{m=1}^N \frac{s_m  \ \sinh( 2 \lambda)}{\sinh (\lambda - \alpha _m)  \sinh (\lambda + \alpha _m) } \sum _{j=1}^M \frac{\sinh(2\lambda)}{\sinh (\lambda - \mu_j) \sinh (\lambda + \mu_j)} \\[1ex]
&+2 \sum _{m,n = 1}^N  \left( s _m s _n + \frac{\sinh (\xi + \alpha _m ) \sinh (\xi - \alpha _m )}{\sinh (\xi + \lambda ) \sinh (\xi - \lambda )} \ s_m \delta _{mn} \right)\times \\[1ex]
& \times 
\left( \coth (\lambda - \alpha _m) \coth (\lambda - \alpha _n) + 2 \coth (\lambda - \alpha _m) \coth (\lambda + \alpha _n) + \coth (\lambda + \alpha _m) \coth (\lambda + \alpha _n) \right) .
\end{split}
\end{equation}

As our next important step toward obtaining the formulas of the algebraic Bethe ansatz for the corresponding Gaudin model we observe that the first term in the expansion of the function $F_M (\mu_1; \mu_2, \ldots, \mu_M)$ \eqref{Be-M} in powers of $\eta$ is  
\begin{equation}
\label{exp-FM}
F_M(\mu_1; \mu_2, \ldots, \mu_M) =  \eta f_M (\mu_1; \mu_2, \ldots, \mu_M)  + \mathcal{O}(\eta ^2) ,
\end{equation}
where
\begin{equation}
\label{f-M}
\begin{split}
f_M (\mu_1; \mu_2, \ldots, \mu_M)  &= \kappa \sinh (2\mu_1) \left( \frac{1}{\sinh (\xi - \mu_1)} - 2 \sinh (\xi + \mu_1) \sum_ {j=2}^M  \frac{1}{\sinh (\mu_1 - \mu_j) \sinh(\mu_1 + \mu_j)} \right. \\[1ex]
&\left. + 2 \sinh (\xi + \mu_1) \sum_ {m=1}^N \frac{s_m}{\sinh (\mu_1 - \alpha_m) \sinh(\mu_1 + \alpha_m )} \right) .
\end{split}
\end{equation}

Along the lines developed in \cite{CAMS,CAMRS,MS2}, we have used the formulas \eqref{Psi1} and \eqref{b1} as well as \eqref{CalB}, \eqref{CalDonOm} and \eqref{deltahat} in order to expand the Bethe vector $\Psi _1(\mu)$ of the XXZ Heisenberg spin chain in powers of $\eta$ and obtained the Bethe vector $\varphi_1 (\mu)$ of the corresponding trigonometric Gaudin model 
\begin{equation}
\label{exp-Psi1}
\Psi _1 (\mu) = \eta \ \varphi _1(\mu) + \mathcal{O}(\eta ^2) ,
\end{equation}
where
\begin{equation}
\label{phi1}
\!\!\!\!\!\!\!\!\!
\varphi _1(\mu) = \kappa \sinh(2\mu) \left( \sum _{m=1}^N  \frac{\sinh(\xi - \alpha_m) \ S^-_m}{\sinh(\mu - \alpha_m) \sinh(\mu + \alpha_m)} + \frac{\psi}{\kappa} \left(  1 +
\sum _{m=1}^N  s_m \  \frac{e^{-2\xi} + \sinh(2\alpha _m) - \cosh(2\mu)}{\sinh(\mu - \alpha_m) \sinh(\mu + \alpha_m)} \right) \right) \Omega _+ .
\end{equation}
The off-shell action of the difference of the transfer matrix of the XXX chain and the central element, the so-called Sklyanin determinant, on the Bethe vector $\Psi _1 (\mu)$ \eqref{Psi1} is obtained from \eqref{Del-calT} and \eqref{t-on-Psi1-tilde} as follows
\begin{equation}
\label{diff-on-Psi1}
\begin{split}
\left( t (\lambda) - \frac{\Delta \left[\mathcal{T}(\lambda) \right] }{\sinh (2\lambda)}  \right) \Psi _1 (\mu) &= \left( \Lambda _1 (\lambda, \mu) - \alpha (\lambda + \eta / 2) \, \widehat{\delta} (\lambda - \eta / 2) \right)  \Psi _1 (\mu) \\
&+ \kappa \sinh ( \xi - \mu ) \frac{\sinh (\eta) \sinh ( 2 (\lambda + \eta) )}{\sinh (\lambda - \mu) \sinh(\lambda + \mu + \eta)} F_1 (\mu) \Psi_1 (\lambda) .
\end{split}
\end{equation}
Finally, the off-shell action of the generating function the Gaudin Hamiltonians on the vector $\varphi_1 (\mu)$ can be obtained from the equation above by using the expansion \eqref{exp-difference} and \eqref{exp-Psi1} on the left hand side as well as the expansion \eqref{exp-LambdaM}, \eqref{exp-FM} and \eqref{exp-Psi1} 
on the right hand side 
\begin{equation}
\label{tau-phi1} 
\tau (\lambda ) \varphi _1(\mu) = \chi _1 (\lambda, \mu) \varphi _1(\mu) + \frac{2}{\kappa} \ \frac{\sinh (\xi - \mu)}{\sinh (\xi - \lambda) \sinh (\xi + \lambda)} \  \frac{\sinh (2\lambda)}{\sinh (\lambda - \mu) \sinh (\lambda + \mu)} \ f_1 (\mu) \varphi _1(\lambda) .
\end{equation}
Therefore $\varphi_1 (\mu)$ \eqref{phi1} is the Bethe vector of the corresponding Gaudin model, i.e.  the eigenvector of the generating function the Gaudin Hamiltonians, with the eigenvalue $ \chi _1 (\lambda, \mu)$ \eqref{chiM}, once the unwanted term is canceled by imposing the corresponding Bethe equation
\begin{equation}
\label{Gm-Be1}
 f_1 (\mu)  = \kappa \sinh (2\mu) \left( \frac{1}{\sinh (\xi - \mu)} + 2 \sinh (\xi + \mu) \sum_ {m=1}^N \frac{s_m}{\sinh (\mu - \alpha_m) \sinh(\mu + \alpha_m )} \right) = 0.
\end{equation}

To obtain the Bethe vector $\varphi_2 (\mu_1,\mu_2)$ of the Gaudin model and the action of the generating function $\tau (\lambda)$ of the Gaudin Hamiltonians on $\varphi_2 (\mu_1,\mu_2)$ we basically follow the steps we have done when studying the action of $\tau (\lambda)$ on $\varphi_1 (\mu)$. The first term in the expansion of the Bethe vector $\Psi _2 ( \mu_1 ,  \mu_2)$ \eqref{Psi2} in powers of $\eta$ yields the corresponding Bethe vector of the Gaudin model
\begin{equation}
\label{Psi2-exp}
\Psi _2(\mu _1 , \mu_2 ) = \eta ^2 \varphi_2 (\mu_1,\mu_2)  + \mathcal{O}(\eta ^3) , 
\end{equation}
where
\begin{align}
\label{phi2}
&\varphi _2 (\mu _1, \mu_2) = \kappa^2 \sinh (2\mu_1) \sinh (2\mu_2) \left( \sum _{m,n=1}^N \frac{\sinh(\xi - \alpha_m) \ S^-_m}{\sinh(\mu_1 - \alpha_m) \sinh(\mu_1 + \alpha_m)}  \ \frac{\sinh(\xi - \alpha_n) \ S^-_n}{\sinh(\mu_2 - \alpha_n) \sinh(\mu_2 + \alpha_n)} \right. \notag \\[1ex]
&+ \frac{\psi}{\kappa} \sum _{m=1}^N \frac{\sinh(\xi - \alpha_m) \ S^-_m}{\sinh(\mu_2 - \alpha_m) \sinh(\mu_2 + \alpha_m)} 
\left( 1 + \sum _{n=1}^N \frac{e^{-2\xi} + \sinh (2\alpha_n) - \cosh(2\mu _1)}{\sinh(\mu _1 - \alpha_n) \sinh(\mu _1 + \alpha_n)} \left( s_n - \delta_{mn} \right) \right) \notag \\[1ex]
&+ \frac{\psi}{\kappa}\sum _{m=1}^N \frac{\sinh(\xi - \alpha_m) \ S^-_m}{\sinh(\mu_1 - \alpha_m) \sinh(\mu_1 + \alpha_m)} \left( 3 + \sum _{n=1}^N \frac{e^{-2\xi} + \sinh (2\alpha_n) - \cosh(2\mu_2)}{\sinh(\mu _2- \alpha_n) \sinh(\mu_2 + \alpha_n)} \ s_n \right) \notag \\[1ex]
&+ e^{-2\xi} \frac{\psi ^2}{\kappa ^2} \sum_{m=1}^N \frac{-e^{\xi - \alpha_m} \cosh(2\mu_1) + \cosh(\xi+\alpha_m)}{\sinh (\mu_1 - \alpha_m) \sinh (\mu_1 + \alpha_m)} \frac{\sinh(\xi - \alpha_m)}{\sinh  (\mu_2 - \alpha_m) \sinh (\mu_2 + \alpha_m)} (2s_m)\notag \\[1ex]
&\left. + \frac{\psi ^2}{\kappa ^2} \left( 1 + \sum _{n=1}^N \frac{e^{-2\xi} + \sinh (2\alpha_m) - \cosh(2\mu_1)}{\sinh(\mu_1 - \alpha_m) \sinh(\mu_1 + \alpha_m)} s_m  \right) 
\left( 3 + \sum _{n=1}^N \frac{e^{-2\xi} + \sinh (2\alpha_n) - \cosh(2\mu_2)}{\sinh(\mu_2 - \alpha_n) \sinh(\mu_2 + \alpha_n)} s_n \right)
\right) \Omega_+ .
\end{align}
Expressing Gaudin Bethe vectors by using creation operators is in accordance with the results in the rational case \cite{CAMS}. There the creation operator was introduced (cf. formula (6.32) in \cite{CAMS}), but here it is necessary to define the family of operators
\begin{equation}
\label{c-K}
\begin{split}
c_K(\mu) &= \kappa \sinh(2\mu) \left(  \sum _{m=1}^N  \frac{\sinh(\xi - \alpha_m) \ S^-_m}{\sinh(\mu - \alpha_m) \sinh(\mu + \alpha_m)} + \frac{\psi}{\kappa} \left(  (-1 + 2K)  \right. \right. \\[1ex]
& \left. \left. +
\sum _{m=1}^N \frac{e^{-2\xi} + \sinh(2\alpha _m) - \cosh(2\mu)}{\sinh(\mu - \alpha_m) \sinh(\mu + \alpha_m)}  \ S^3_m   \right) + e^{-2\xi} \frac{\psi^2}{\kappa ^2} \sum _{m=1}^N  \frac{\cosh(\xi + \alpha_m) - e^{\xi - \alpha_m} \cosh(2\mu)}{\sinh(\mu - \alpha_m) \sinh(\mu + \alpha_m)} \ S^+_m \right) ,
\end{split}
\end{equation}
for any natural number $K$. Thus the Bethe vectors \eqref{phi1} and \eqref{phi2} can be expressed as
\begin{equation}
\label{phi1,2-c}
\varphi _1 (\mu ) = c_1(\mu) \Omega _+ \qquad \text{and} \qquad
\varphi _2 (\mu _1, \mu_2) = c_1(\mu _1) c_2(\mu_2) \Omega _+ .
\end{equation}
Although in general the operators \eqref{c-K} do not commute, it is straightforward to check that
the Bethe vector $\varphi _2 (\mu _1, \mu_2)$ is a symmetric function
\begin{equation}
\label{phi2-sym}
\varphi _2 (\mu _1, \mu_2) = c_1(\mu _1) c_2(\mu_2) \Omega _+ = c_1(\mu _2) c_2(\mu_1) \Omega _+ = \varphi _2 (\mu _2, \mu_1) . 
\end{equation}
 
It is of interest to study the action of the difference of the transfer matrix $t(\lambda)$ and the so-called Sklyanin determinant $\Delta \left[\mathcal{T}(\lambda)\right]$ on the Bethe vector $\Psi _2(\mu _1 , \mu_2 )$ using \eqref{Del-calT} and \eqref{t-on-Psi2-tilde}
\begin{align}
\label{diff-on-Psi2}
\left( t (\lambda) - \frac{\Delta \left[\mathcal{T}(\lambda) \right] }{\sinh (2\lambda)}  \right) \Psi _2 (\mu_1 , \mu_2) &=  \left( \Lambda _2 (\lambda , \{\mu _i \}_{i=1}^2 ) - \alpha (\lambda + \eta / 2) \, \widehat{\delta} (\lambda - \eta / 2) \right) \Psi _2 (\mu_1 , \mu_2) \notag \\[1ex]
&+ \frac{ \sinh (\eta) \sinh (2 (\lambda + \eta ))}{\sinh (\lambda - \mu _1) \sinh (\lambda + \mu _1 + \eta )} \kappa \sinh (\xi - \mu _1 ) \ F_2(\mu _1; \mu _2) \Psi _2 ( \lambda ,  \mu_2) \notag \\[1ex]
&+ \frac{ \sinh (\eta) \sinh (2 (\lambda + \eta ))}{\sinh (\lambda - \mu _2) \sinh (\lambda + \mu _2 + \eta )} \kappa \sinh (\xi - \mu _2 ) \ F_2(\mu _2; \mu _1) \Psi _2 ( \lambda ,  \mu_1) .
\end{align}
The off-shell action of the generating function of the Gaudin Hamiltonians on the Bethe vector $\varphi_2 (\mu_1,\mu_2)$ is obtained from the equation above using the expansions \eqref{exp-difference} and \eqref{Psi2-exp} on the left hand side and \eqref{exp-LambdaM}, \eqref{Psi2-exp} and 
\eqref{exp-FM} on the right hand side. Then, by comparing the terms of the fourth power in $\eta$ on both sides of \eqref{diff-on-Psi2} we obtain  
\begin{equation}
\label{tau-phi2}
\begin{split}
\tau (\lambda) \varphi_2 (\mu_1, \mu_2) &= \chi _2 (\lambda , \mu _1, \mu _2) \varphi_2 (\mu_1,\mu_2)   + \frac{2}{\kappa} \ \frac{\sinh (2\lambda)}{\sinh (\xi - \lambda) \sinh (\xi + \lambda)}  \times \\[1ex]
& \times \left( \frac{\sinh (\xi - \mu_1)}{\sinh (\lambda - \mu_1) \sinh (\lambda + \mu_1)}   f_2 (\mu_1;\mu_2)  \varphi_2 (\lambda, \mu_2) \right. \\[1ex]
&\left. +  \frac{\sinh (\xi - \mu_2)}{\sinh (\lambda - \mu_2) \sinh (\lambda + \mu_2)} f_2 (\mu_2; \mu_1) \varphi_2 (\lambda, \mu_1) \right).
\end{split}
\end{equation}
The two unwanted terms on the right hand side of the equation above are annihilated by the following Bethe equations 
\begin{align}
\label{Gm-Be2-1}
f_2 (\mu_1; \mu_2)  &= \kappa \sinh (2\mu_1) \left( \frac{1}{\sinh (\xi - \mu_1)} -  \frac{2 \sinh (\xi + \mu_1)}{\sinh (\mu_1 - \mu_2) \sinh(\mu_1 + \mu_2)} \right. \notag \\
&\left. + 2 \sinh (\xi + \mu_1) \sum_ {m=1}^N \frac{s_m}{\sinh (\mu_1 - \alpha_m) \sinh(\mu_1 + \alpha_m )} \right) = 0 , \\[1ex]
\label{Gm-Be2-2}
f_2 (\mu_2; \mu_1)  &= \kappa \sinh (2\mu_2) \left( \frac{1}{\sinh (\xi - \mu_2)} -  \frac{2 \sinh (\xi + \mu_2)}{\sinh (\mu_2 - \mu_1) \sinh(\mu_2 + \mu_1)} \right. \notag \\
&\left. + 2 \sinh (\xi + \mu_2) \sum_ {m=1}^N \frac{s_m}{\sinh (\mu_2 - \alpha_m) \sinh(\mu_2 + \alpha_m )} \right) = 0 .
\end{align}

In general, we have that the first term in the expansion of the Bethe vector \break $\Psi _M ( \mu_1 ,  \mu_2 ,  \dots ,  \mu_M )$ \eqref{PsiM} in powers of $\eta$ is
\begin{equation}
\label{exp-PsiM}
\Psi _M ( \mu_1 ,  \mu_2 ,  \dots ,  \mu_M ) = \eta ^M \varphi_M ( \mu_1 ,  \mu_2 ,  \dots ,  \mu_M )  + \mathcal{O}(\eta ^{M+1}) , 
\end{equation}
where $M$ is a natural number and
\begin{equation}
\label{phiM}
\varphi_M ( \mu_1 ,  \mu_2 ,  \dots ,  \mu_M ) = c_1 (\mu _1) c_2 (\mu _2) \cdots c_M (\mu _M)  \Omega _+ ,
\end{equation}
and the operator $c_K(\mu_K)$, $K=1, \ldots , M$, are given in \eqref{c-K}.

Although the operators $c_K(\mu_K)$ do not commute, the Bethe vector of the Gaudin model $\varphi_M ( \mu_1 ,  \mu_2 ,  \dots ,  \mu_M )$ is a symmetric function of its arguments, since a straightforward calculation shows that the operators $c_K(\mu)$  satisfy the following identity,
\begin{equation}
\label{F-commute}
c_K(\mu) c_{K+1}(\mu^{\prime}) - c_K(\mu^{\prime}) c_{K+1}(\mu) = 0,
\end{equation}
for $K=1, \ldots , M-1$.  The action of the generating function $\tau (\lambda)$ \eqref{b-tau} on the Bethe vector $\varphi_M ( \mu_1 ,  \mu_2 ,  \dots ,  \mu_M )$ can be derived as in the two previous  cases when $M = 1$ \eqref{tau-phi1} and $M = 2$ \eqref{tau-phi2}. In the present case we use the expansions 
\eqref{exp-LambdaM}, \eqref{exp-FM} and \eqref{exp-PsiM} to obtain 
\begin{equation}
\label{tau-on-phiM}
\begin{split}
\tau (\lambda) \varphi_M ( \mu_1 ,  \mu_2 ,  \dots ,  \mu_M ) &= \chi _M(\lambda , \{\mu _i \} _{i=1}^M) \varphi_M ( \mu_1 ,  \mu_2 ,  \dots ,  \mu_M ) + \frac{2}{\kappa} \ \frac{\sinh (2\lambda)}{\sinh (\xi - \lambda) \sinh (\xi + \lambda)} \times \\
&\times \sum _{i=1}^M \frac{\sinh (\xi - \mu_i)}{\sinh (\lambda - \mu_i) \sinh (\lambda + \mu_i)}
f_M (\mu_i; \{\mu _j \} _{j \neq i} ) \varphi_M ( \lambda ,  \{\mu _j \} _{j \neq i} ) ,
\end{split}
\end{equation}
where $\chi _M(\lambda , \{\mu _i \} _{i=1}^M)$ is given in \eqref{chiM} and the unwanted terms on the right hand side of the equation above are canceled by the following Bethe equations 
\begin{equation}
\label{GM-BEM-i}
\begin{split}
f_M (\mu_i; \{\mu _j \} _{j \neq i}^M )  &= \kappa \sinh (2\mu_i) \left( \frac{1}{\sinh (\xi - \mu_i)} - 2 \sinh (\xi + \mu_i) \sum_ {j=2}^M  \frac{1}{\sinh (\mu_i - \mu_j) \sinh(\mu_i + \mu_j)} \right. \\[1ex]
&\left. + 2 \sinh (\xi + \mu_i) \sum_ {m=1}^N \frac{s_m}{\sinh (\mu_i - \alpha_m) \sinh(\mu_i + \alpha_m )} \right) = 0,
\end{split}
\end{equation}
for $i = 1 , 2 , \dots M$. As expected, due to our definition of the Bethe vector $\varphi_M ( \mu_1 ,  \mu_2 ,  \dots ,  \mu_M )$ \eqref{phiM}, the quasi-classical limit has yielded the above simple formulae for the off-shell action of the generating function $\tau (\lambda)$. 

An alternative approach to the implementation of the algebraic Bethe ansatz for the trigonometric $s\ell(2)$ Gaudin model, with the triangular K-matrix \eqref{K-min-Gm}, is based on the corresponding non-unitary classical r-matrix. This study will be reported in \cite{MS2}.

\section{Conclusions \label{sec: Conclu}}
We have implemented fully the off-shell algebraic Bethe ansatz for the XXZ Heisenberg spin chain 
in the case when both boundary matrices have the upper-triangular form.  As opposed to the case of the XXX Heisenberg spin chain where the general reflection matrices could be put into the upper triangular form without any loss of generality \cite{Eric13, CAMS}, here the triangularity of the reflection matrices has to be imposed as extra conditions on the respective parameters. A suitable realization for the Sklyanin monodromy matrix is obtained as a direct consequence of the identity satisfied by the Lax operator.  This realization led to the action of the entries of the Sklyanin monodromy matrix on the vector $\Omega _+$ and consequently to the observation that $\Omega _+$ is an eigenvector of the transfer matrix of the chain.

The essential step of the algebraic Bethe ansatz is the definition of the corresponding Bethe vectors. Initially we have obtained the Bethe vectors $\widetilde{\Psi} _M ( \mu_1 ,  \mu_2 ,  \dots ,  \mu_M )$, for $M = 1, 2, 3, 4$, by requiring that their scaling limit corresponds to the Bethe vectors of the XXX Heisenberg chain. We gave a step by step presentation of the $M = 1, 2, 3$ Bethe vectors, including the formulae for the action of $t(\lambda)$, the corresponding eigenvalues and Bethe equations. In this way we have exposed the property of these vectors to make the off shell action of the transform matrix as simple as possible. We did not present here all the necessary formulae of the Bethe vector $\widetilde{\Psi }_4 ( \mu_1,  \mu_2, \mu_3, \mu_4 )$, as they are cumbersome. More importantly, they do not admit any compact closed form for an arbitrary natural number M. However, we have noticed the identities  \eqref{identity-PsiM-Li} and \eqref{identity-PsiM-Fi,j} which enabled the general form of the Bethe vectors for a fixed $M$. The general form of Bethe vectors can be expressed as a sum of a particular one and a linear combination of lower order Bethe vectors that correspond to the same eigenvalue \eqref{general-Psi}. This is indeed the case with Bethe vectors of any order, for details see Appendix \ref{app: B-vec}. A careful analysis reveals that there exists a particular form of the Bethe vector $\Psi _M ( \mu_1 ,  \mu_2 ,  \dots ,  \mu_M )$ which, for an arbitrary natural number $M$, can be defined by the suitable recurrence procedure analogous to the one proposed in the case of the XXX Heisenberg chain \cite{CAMS}. Actually, the recurrence relations defining the relevant coefficient functions differ only in the multiplicative factors from the respective ones in the case of the XXX Heisenberg chain. As expected, the action of $t(\lambda)$ on the Bethe vector $\Psi _M ( \mu_1 ,  \mu_2 ,  \dots ,  \mu_M )$ is again very simple. Actually, the action of the transfer matrix is as simple as it could possible be since it almost coincides with the corresponding action in the case when the two boundary matrices are diagonal \cite{Sklyanin88, Hikami95}. 

As in the case of the XXX Heisenberg chain \cite{CAMRS}, the quasi-classical expansion of the linear combination of the transfer matrix of the XXZ Heisenberg spin chain and the central element, the so-called Sklyanin determinant yields the generating function of the trigonometric Gaudin Hamiltonians with boundary terms \cite{MS2}. Based on this result, and the appropriate definition of the corresponding Bethe vectors $\varphi_M ( \mu_1 ,  \mu_2 ,  \dots ,  \mu_M )$, we showed how the quasi-classical limit yields the off-shell action of the generating function of the Gaudin Hamiltonians as well as the spectrum and the Bethe equations. As opposed to the rational case where the Gaudin Bethe vectors were defined by the action of the creation operator \cite{CAMS}, here it was necessary to define the family of operators. As in the case of the spin chain, the off-shell action of the generating function $\tau (\lambda)$ on the Bethe vectors $\varphi_M ( \mu_1 ,  \mu_2 ,  \dots ,  \mu_M )$ is strikingly simple.  It is as simple as it can be since it practically coincide with the corresponding formula in the case when the boundary matrix is diagonal \cite{Hikami95}. 

It would be of interest to establish a relation between Bethe vectors of the Gaudin model and solutions to the corresponding generalized Knizhnik-Zamolodchikov equations, along the lines it was done in the case  when the boundary matrix is diagonal \cite{Hikami95},  as well as to study possible relations between Bethe vectors of XXZ chain obtained in the Section \ref{sec: ABA-XXZchain} and the solutions to the boundary quantum Knizhnik-Zamolodchikov equations \cite{Jimbo95b, Reshetikhin16, Reshetikhin16b}.

\bigskip

\noindent
\textbf{Acknowledgments}

\noindent
We acknowledge partial financial support by the FCT project PTDC/MAT-GEO/3319/2014.
I.S. was supported in part by the Serbian Ministry of Science and Technological Development under grant number ON 171031.

\appendix
\section{Basic definitions \label{app: basic-def}}
We consider a spin chain with N sites with spin $s$ representations, i.e. a local $\mathbb{C}^{2s+1}$ space at each site and the operators 
\begin{equation}
S_m^{\alpha} = \mathbbm{1} \otimes \cdots \otimes \underbrace{S^{\alpha}} _m \otimes \cdots \otimes \mathbbm{1},
\end{equation}
with $\alpha = +,-, 3$ and $m= 1, 2 ,\dots , N$. The operators $S^{\alpha}$ with $\alpha = +, - , 3$, act in some (spin $s$) representation space $\mathbb{C}^{2s+1}$ with the commutation relations \cite{KulishResh83,FRT89,Anastasia07}
\begin{equation}
\label{crspin-s}
[S^3, S^{\pm}] = \pm S^{\pm}, \quad [S^+,S^-] =\frac{\sinh (2\eta S^3)}{\sinh (\eta)} = [2 S^3]_q ,
\end{equation}
with $q = e^{\eta}$ , 
and Casimir operator
\begin{equation}
\label{Casimir}
c_2 = \frac{q+q^{-1}}{2}[S^3]_q ^2 + \frac{1}{2} (S^+S^-+S^-S^+) = \frac{q+q^{-1}}{2}[S^3]_q ^2  
+ \frac{1}{2} [2S^3]_q + S^-S^+ .
\end{equation}
In the space $\mathbb{C}^{2s+1}$ these operators admit the following matrix representation 
\begin{equation}
\label{matrix-rep}
S^{3} = \sum _{i=1}^{2s+1} a_i e_{i\, i} , \quad S^{+} = \sum _{i=1}^{2s+1} b_i e_{i\, i+1} ,
\quad S^{-} = \sum _{i=1}^{2s+1} b_i e_{i+1\, i}  \quad \text{and} \quad c_2  = [s+1]_q \, [s]_q \ \mathbbm{1} ,
\end{equation}
where
\begin{equation}
\label{matrix-elements}
(e_{ij})_{kl} = \delta _{i\,k} \delta _{j\, l}, \quad a_i  = s+1 - i , \quad
b_i = \sqrt{[i]_q \ [2s+1 -i]_q}  \quad \text{and} \quad  [x]_q = \frac{q^x - q^{-x}}{q - q^{-1}}.
\end{equation}
In the particular case of spin $\frac12$ representation, one recovers the Pauli matrices
$$
S^{\alpha} = \frac{1}{2} \sigma ^{\alpha} = \frac{1}{2} \left(\begin{array}{cc}
\delta_{\alpha3} & 2\delta_{\alpha+}  \\
2\delta_{\alpha-} & - \delta _{\alpha 3} \end{array}\right).
$$

\section{Commutation relations \label{app: commut-rel}}
The equation \eqref{RE-algebra} yields the exchange relations between the operators $\mathcal{A} (\lambda)$, $\mathcal{B} (\lambda)$, $\mathcal{C} (\lambda)$  and $\widehat{\mathcal{D}} (\lambda)$.
The relevant relations are
\begin{align}
\label{comm-relBB+CC}
\mathcal{B} (\lambda) \mathcal{B} (\mu) &= \mathcal{B} (\mu) \mathcal{B} (\lambda) , \qquad  
\mathcal{C} (\lambda) \mathcal{C} (\mu) = \mathcal{C} (\mu) \mathcal{C} (\lambda) ,  \\[2ex]
\label{comm-relAB}
\mathcal{A} (\lambda) \mathcal{B} (\mu) &= 
\frac{\sinh(\lambda + \mu)\sinh(\lambda - \mu - \eta)}{\sinh(\lambda - \mu)\sinh(\lambda + \mu + \eta)} \mathcal{B} (\mu) \mathcal{A} (\lambda) 
+ \frac{\sinh(\eta) \sinh(2\mu)}{\sinh(\lambda - \mu)\sinh(2 \mu + \eta)} \mathcal{B} (\lambda) \mathcal{A} (\mu) \notag \\
&- \frac{\sinh(\eta)}{\sinh(\lambda +\mu + \eta)} \mathcal{B} (\lambda) \widehat{\mathcal{D}} (\mu) , \\[2ex]
\label{comm-rel-hDB}
\widehat{\mathcal{D}} (\lambda) \mathcal{B} (\mu) &= \frac{\sinh(\lambda - \mu + \eta)\sinh(\lambda + \mu + 2 \eta)}{\sinh(\lambda - \mu) \sinh(\lambda + \mu + \eta)} \mathcal{B} (\mu) \widehat{\mathcal{D}} (\lambda) 
- \frac{\sinh(\eta) \sinh(2(\lambda + \eta))} {\sinh(\lambda - \mu)\sinh(2 \lambda + \eta)} \mathcal{B} (\lambda) \widehat{\mathcal{D}} (\mu) \notag \\
&+ \frac{ \sinh(\eta) \sinh(2\mu) \sinh(2(\lambda + \eta))}{\sinh(2\lambda + \eta) \sinh(2\mu + \eta)\sinh(\lambda +\mu + \eta)} \mathcal{B} (\lambda) \mathcal{A} (\mu) , \\[2ex]
\label{comm-relCB}
\left[ \mathcal{C} (\lambda)  , \mathcal{B} (\mu) \right] &= \frac{\sinh (\eta) \sinh (2\lambda) \sinh(\lambda - \mu + \eta)}{\sinh(\lambda - \mu) \sinh(2 \lambda + \eta) \sinh(\lambda + \mu + \eta) } \mathcal{A} (\mu) \mathcal{A} (\lambda) \notag \\
&- \frac{\sinh^2(\eta)  \sinh (2\lambda)}{\sinh(\lambda - \mu) \sinh(2 \lambda + \eta) \sinh(2 \mu + \eta)} \mathcal{A} (\lambda) \mathcal{A} (\mu)  \notag \\
&+\frac{\sinh(\eta) \sinh(\lambda + \mu) }{\sinh(\lambda-\mu)\sinh(\lambda + \mu + \eta)} \mathcal{A} (\mu) \widehat{\mathcal{D}} (\lambda)  
- \frac{\sinh(\eta) \sinh(2\lambda)}{\sinh(\lambda - \mu) \sinh(2 \lambda + \eta)} \mathcal{A} (\lambda) \widehat{\mathcal{D}} (\mu) \notag \\
&-\frac{\sinh ^2 (\eta)}{\sinh(2 \mu + \eta) \sinh(\lambda + \mu + \eta) }\widehat{\mathcal{D}} (\lambda) \mathcal{A} (\mu)
- \frac{\sinh(\eta)}{\sinh(\lambda + \mu + \eta)} \widehat{\mathcal{D}} (\lambda) \widehat{\mathcal{D}} (\mu) .
\end{align}
For completeness we include the following commutation relations 
\begin{align}
\label{AA}
\left [ \mathcal{A} (\lambda) , \mathcal{A} (\mu) \right] &= \frac{\sinh(\eta)}{\sinh(\lambda + \mu + \eta)}  \left( \mathcal{B} (\mu) \mathcal{C} (\lambda) - \mathcal{B} (\lambda) \mathcal{C} (\mu) \right) \\[2ex]
\label{AD}
\left [ \mathcal{A} (\lambda) , \widehat{\mathcal{D}} (\mu)  \right] &= \frac{\sinh(\eta) \sinh(2(\mu + \eta))}{\sinh(\lambda - \mu) \sinh(2\mu + \eta)} \left( \mathcal{B} (\lambda) \mathcal{C} (\mu) - \mathcal{B} (\mu) \mathcal{C} (\lambda) \right) \\[2ex]
\label{DD}
\left [ \widehat{\mathcal{D}} (\lambda) , \widehat{\mathcal{D}} (\mu)  \right] &= \frac{\sinh(\eta) \sinh(2(\lambda + \eta)) \sinh(2(\mu + \eta))}{\sinh(2\lambda + \eta)\sinh(2\mu + \eta) \sinh(\lambda + \mu + \eta)} \left( \mathcal{B} (\lambda) \mathcal{C} (\mu) - \mathcal{B} (\mu) \mathcal{C} (\lambda) \right)
\end{align}
The implementation of the algebraic Bethe ansatz presented in Section \ref{sec: ABA-XXZchain} is based
on the above relations. For convenience, we also include the following three relations which follow from the ones above and are essential in the derivation of the off-shell action \eqref{t-on-Psi2-til} of the transfer matrix of the inhomogeneous XXZ chain \eqref{transfer-matrix} on the Bethe vector $\widetilde{\Psi} _2 ( \mu_1 ,  \mu_2)$ \eqref{Psi2-til}
\begin{equation}
\label{ABBOmega}
\begin{split}
\mathcal{A} (\lambda) \mathcal{B} (\mu_1) \mathcal{B} (\mu_2) \Omega _+
&= \prod _{i = 1}^2
\frac{\sinh (\lambda + \mu_i) \sinh(\lambda - \mu_i - \eta)}{\sinh (\lambda - \mu _i) \sinh(\lambda + \mu_i + \eta)} 
\alpha (\lambda) \mathcal{B} (\mu_1) \mathcal{B} (\mu_2)  \Omega _+ \\
&+ \sum _{i=1} ^2 \frac{\sinh (\eta) \sinh (2 \mu_i)}{\sinh (2 \mu _i + \eta) \sinh (\lambda - \mu_i)} \frac{\sinh (\mu_i + \mu_{3-i}) \sinh(\mu_i  - \mu_{3-i} - \eta)}{\sinh (\mu_i - \mu _{3-i}) \sinh(\mu_i + \mu_{3-i} + \eta)} \times \\
&\times \alpha (\mu_i) \mathcal{B} (\lambda) \mathcal{B} (\mu_{3-i})  \Omega _+ \\
&- \sum _{i=1} ^2 \frac{\sinh (\eta)}{\sinh (\lambda + \mu_i + \eta)} \frac{\sinh (\mu_i - \mu_{3-i} + \eta)\sinh (\mu_i + \mu_{3-i} + 2\eta)}{\sinh (\mu_i - \mu _{3-i}) \sinh(\mu_i + \mu_{3-i} + \eta)}  \times \\
&\times \widehat{ \delta} (\mu_i) \mathcal{B} (\lambda) \mathcal{B} (\mu_{3-i}) \Omega _+  ,
\end{split}
\end{equation}
analogously, 
\begin{equation}
\label{DBBOmega}
\begin{split}
\widehat{\mathcal{D}} (\lambda) \mathcal{B} (\mu_1) \mathcal{B} (\mu_2) \Omega _+
&= \prod _{i = 1}^2
\frac{\sinh (\lambda - \mu_i + \eta) \sinh(\lambda + \mu_i + 2 \eta )}{\sinh (\lambda - \mu_i ) \sinh (\lambda + \mu _i + \eta)}  
\widehat{ \delta} (\lambda) \mathcal{B} (\mu_1) \mathcal{B} (\mu_2)  \Omega _+ \\
&-  \sum _{i=1} ^2 \frac{\sinh (\eta) \sinh (2 (\lambda + \eta ))}{\sinh (2 \lambda + \eta) \sinh (\lambda - \mu_i)} \frac{\sinh (\mu_i  - \mu_{3-i} + \eta) \sinh (\mu_i + \mu_{3-i} + 2 \eta)}{\sinh (\mu_i - \mu _{3-i})\sinh (\mu_1 + \mu_{3-i} + \eta)} \times \\
& \times \widehat{ \delta} (\mu_i) \mathcal{B} (\lambda) \mathcal{B} (\mu_{3-i}) \Omega _+ \\
&+ \sum _{i=1} ^2 \frac{\sinh (\eta) \sinh (2 \mu_i)  \sinh (2(\lambda + \eta))}{\sinh (2 \lambda + \eta) \sinh (2 \mu _i + \eta) \sinh (\lambda + \mu_i + \eta)} \times \\
& \times 
\frac{\sinh (\mu_i + \mu_{3-i} ) \sinh (\mu_i - \mu_{3-i} - \eta)}{\sinh (\mu_i - \mu _{3-i}) \sinh (\mu_i + \mu_{3-i} + \eta)} \alpha (\mu_i) \mathcal{B} (\lambda) \mathcal{B} (\mu_{3-i}) \Omega _+ ,
\end{split}
\end{equation}
and finally, 
\begin{align}
\label{CBBOmega}
&\mathcal{C} (\lambda) \mathcal{B} (\mu_1) \mathcal{B} (\mu_2)\Omega _+ =  \sum _{i=1}^2
\left( 
\frac{ \sinh (\eta)  \sinh (2 \mu_i ) \sinh (2 \lambda) }{\sinh (2 \lambda + \eta) \sinh (2 \mu_i + \eta) \sinh (\lambda + \mu _i + \eta)} \times 
\right. \notag \\
&\times \frac{\sinh (\lambda + \mu_{3-i}) \sinh (\lambda - \mu_{3-i} - \eta)}{\sinh (\lambda - \mu_{3-i})\sinh (\lambda + \mu_{3-i} + \eta)} \frac{\sinh (\mu_i + \mu_{3-i}) \sinh(\mu_i - \mu_{3-i} - \eta)}{\sinh (\mu_i - \mu_{3-i}) \sinh(\mu_i + \mu_{3-i} + \eta)} \alpha (\lambda) \alpha (\mu_i) \notag \\
&- \frac{\sinh (\eta) \sinh (2 \lambda)}{\sinh (\lambda -  \mu _i) \sinh (2 \lambda + \eta)} 
\frac{\sinh (\lambda + \mu_{3-i}) \sinh (\lambda - \mu_{3-i} - \eta)}{\sinh (\lambda - \mu_{3-i}) \sinh (\lambda + \mu_{3-i} + \eta)} \times \notag \\
&\times \frac{\sinh (\mu_i - \mu_{3-i} + \eta) \sinh (\mu_i + \mu_{3-i} + 2 \eta)}{\sinh (\mu_i - \mu_{3-i}) \sinh (\mu_i + \mu_{3-i} + \eta)} \alpha (\lambda) \widehat{\delta} (\mu_i) \notag \\
&+ \frac{\sinh (\eta) \sinh (2 \mu_i)}{\sinh (\lambda -  \mu _i) \sinh (2 \mu_i + \eta)} 
\frac{\sinh (\lambda - \mu_{3-i} + \eta) \sinh (\lambda + \mu_{3-i} + 2 \eta)}{\sinh (\lambda - \mu_{3-i}) \sinh (\lambda + \mu_{3-i} + \eta)} \times \notag \\
& \times \frac{\sinh (\mu_i + \mu_{3-i}) \sinh (\mu_i - \mu_{3-i} - \eta)}{\sinh (\mu_i - \mu_{3-i}) 
\sinh (\mu_i + \mu_{3-i} + \eta)} \alpha (\mu_i) \widehat{\delta} (\lambda) \notag \\
&- \frac{\sinh (\eta)}{\sinh (\lambda + \mu _i + \eta)} \frac{\sinh (\lambda - \mu_{3-i} + \eta) \sinh (\lambda + \mu_{3-i} + 2 \eta)}{\sinh (\lambda - \mu_{3-i}) \sinh (\lambda + \mu_{3-i} + \eta)} \times \notag \\
& \times \left. \frac{\sinh (\mu_i - \mu_{3-i} + \eta) \sinh (\mu_i + \mu_{3-i} + 2 \eta)}{\sinh (\mu_i - \mu_{3-i}) \sinh (\mu_i + \mu_{3-i} + \eta)} \widehat{\delta} (\lambda) \widehat{\delta} (\mu_i) \right)
\mathcal{B} (\mu_{3-i}) \Omega _+ \notag \\
&+ \left( \frac{\sinh ^2 (\eta)   \sinh (2 \mu_1)  \sinh (2 \mu_2) \sinh (\mu_1 + \mu_2)
}{\sinh (\lambda - \mu_1) \sinh (\lambda - \mu_2) \sinh (2 \mu_1 + \eta) \sinh (2 \mu_2 + \eta) 
} \right. \times \notag \\
&\times \left. \frac{\sinh (\lambda + \mu_1) \sinh (\lambda - \mu_2 + \eta) + \sinh (\lambda - \mu_1) \sinh (\lambda + \mu_2 + \eta)}{\sinh (\lambda + \mu_1 + \eta)  \sinh (\lambda + \mu_2 + \eta) \sinh (\mu_1 + \mu_2 + \eta)} \alpha (\mu _1) \alpha (\mu _2)  \right. \notag \\
&- \frac{\sinh ^2 (\eta) \sinh (2 \mu_1) \sinh (\mu_1 - \mu_2 - \eta)}{\sinh (\lambda - \mu _1)  
\sinh (\lambda - \mu _2) \sinh (2 \mu _1 + \eta) \sinh (\mu _1 - \mu _2)} \times \notag \\
&\times \frac{\sinh (\lambda - \mu _1) \sinh (\lambda - \mu_2) + \sinh (\lambda + \mu _1) \sinh (\lambda + \mu_2 + 2 \eta )}{\sinh (\lambda + \mu_1 + \eta) \sinh(\lambda + \mu_2 + \eta)}
\alpha (\mu_1)  \widehat{\delta} (\mu_2)  \notag \\
&- \frac{\sinh ^2 (\eta) \sinh (2 \mu_2) \sinh (\mu_2 - \mu_1 - \eta)}{\sinh (\lambda - \mu_1) \sinh (\lambda - \mu_2) \sinh (2 \mu_2 + \eta) \sinh (\mu_2 - \mu_1)} \times \notag \\
&\times \frac{\sinh (\lambda - \mu _1) \sinh (\lambda - \mu_2) + \sinh (\lambda + \mu _1) \sinh (\lambda + \mu_2 + 2 \eta )}{\sinh (\lambda + \mu_1 + \eta) \sinh (\lambda + \mu_2 + \eta)}
\alpha (\mu _2) \widehat{\delta} (\mu_1)  \notag \\
&- \frac{\sinh ^2 (\eta)  \sinh (\mu _1 + \mu_2 + 2 \eta)}{\sinh (\lambda - \mu_1)  \sinh (\lambda - \mu_2)  \sinh (\mu _1 + \mu_2 + \eta) } \times \notag \\
&\times \left. \frac{\sinh (\lambda + \mu _1 + 2 \eta) \sinh (-\lambda + \mu _2 + \eta) + \sinh (-\lambda + \mu _1) \sinh (\lambda + \mu _2 + \eta)}
{\sinh (\lambda + \mu_1 + \eta) \sinh (\lambda + \mu_2 + \eta)}
\widehat{\delta} (\mu_1)  \widehat{\delta} (\mu_2) \right) \mathcal{B} (\lambda) \Omega _+ .
\end{align}

\section{Bethe vectors \label{app: B-vec}} 
With the aim of pursuing the general case in this appendix we present the Bethe vector $\widetilde{\Psi} _3 ( \mu_1 ,  \mu_2, \mu_3)$, which in the scaling limit corresponds to the corresponding Bethe vector of the XXX chain \cite{CAMS}, 
\begin{equation}
\label{Psi3-til}
\begin{split}
\widetilde{\Psi} _3 ( \mu_1 ,  \mu_2 ,  \mu_3) = \mathcal{B} (\mu_1) \mathcal{B} (\mu_2) \mathcal{B} (\mu_3)\Omega _+ 
+  \widetilde{b} ^{(1)}_3(\mu_3 ; \mu_2 , \mu_1) \mathcal{B} (\mu_1) \mathcal{B} (\mu_2) \Omega _+ 
+ \widetilde{b} ^{(1)}_3(\mu_1 ;  \mu_2 ,  \mu_3)  \times \\
\times \mathcal{B} (\mu_2) \mathcal{B} (\mu_3) \Omega _+ 
+ \widetilde{b} ^{(1)}_3(\mu_2 ; \mu_1 , \mu_3) \mathcal{B} (\mu_1) \mathcal{B} (\mu_3) \Omega _+ 
+ \widetilde{b} ^{(2)}_3(\mu_1 ,  \mu_2 ;  \mu_3)  \mathcal{B} (\mu_3) \Omega _+  \\
+ \widetilde{b} ^{(2)}_3(\mu_1, \mu_3 ;  \mu_2) \mathcal{B} (\mu_2) \Omega _+ 
+ \widetilde{b} ^{(2)}_3(\mu_2, \mu_3 ;  \mu_1) \mathcal{B} (\mu_1) \Omega _+ 
+ \widetilde{b} ^{(3)}_3 ( \mu_1 ,  \mu_2 ,  \mu_3) \Omega _+ ,
\end{split}
\end{equation}
where the coefficient functions $\widetilde{b} ^{(1)}_3 (\mu_1 ;  \mu_2 ,  \mu_3)$ , $\widetilde{b} ^{(2)}_3 (\mu_1 ,  \mu_2 ;  \mu_3)$ and $\widetilde{b} ^{(3)}_3 (\mu_1 ,  \mu_2 ,  \mu_3)$ are explicitly given  by
\begin{align}
\label{1b3-tilde}
\widetilde{b} ^{(1)}_3 (\mu_1 ;  \mu_2 ,  \mu_3) &= \left( - \frac{\psi^+}{\kappa ^+} \right)
 \left( \prod _{i=2}^3 \frac{\sinh (\mu _1 + \mu _i) \sinh (\mu _1 - \mu _i - \eta )}{\sinh (\mu _1 - \mu _i) \sinh (\mu _1 + \mu _i + \eta )} \ \frac{ \sinh (2 \mu _1)}{\sinh (2 \mu _1  + \eta)}
\ \cosh (\xi ^+ - \mu _1) \ \alpha (\mu _1) \right. \notag \\[1ex]
&\left. - \prod _{i=2}^3 \frac{\sinh (\mu _1 - \mu_i + \eta) \sinh (\mu _1 + \mu_i + 2 \eta)}{\sinh (\mu _1 - \mu _i) \sinh (\mu _1 + \mu_i + \eta)} \ \cosh (\xi ^+ + \mu _1 + \eta)  \ \widehat{ \delta}  (\mu _1) \right) , 
\end{align}
\begin{align}
\label{2b3-tilde}
&\widetilde{b} ^{(2)}_3 (\mu_1 ,  \mu_2 ;  \mu_3) = \left( \frac{\psi^+}{\kappa ^+} \right) ^2 \left( \prod _{i=1}^2 \frac{\sinh (2 \mu _i)}{\sinh (2 \mu _i + \eta)} \ \frac{\sinh (\mu _i + \mu _3) \sinh (\mu _i - \mu _3 - \eta )}{\sinh (\mu _i - \mu _3) \sinh (\mu _i + \mu _3 + \eta )} \
\frac{\sinh (\mu _1 + \mu_2)}{\sinh (\mu _1 + \mu _2 + \eta)} \times \right. \notag \\[1ex]
&  \times \cosh (2 \xi ^+ - \mu _1 - \mu _2 + \eta ) \ \alpha (\mu _1) \alpha (\mu _2) - \frac{\sinh (2 \mu _1)}{\sinh (2 \mu _1 + \eta)} \ \frac{\sinh (\mu _1 + \mu _3) \sinh (\mu _1 - \mu _3 - \eta )}{\sinh (\mu _1 - \mu _3) \sinh (\mu _1 + \mu _3 + \eta )} \  \times \notag \\[1ex]
&\times  \frac{\sinh (\mu _2 - \mu_3 + \eta) \sinh (\mu _2 + \mu_3 + 2 \eta)}{\sinh (\mu _2 - \mu _3) \sinh (\mu _2 + \mu_3 + \eta)} \ \frac{\sinh (\mu_1 - \mu_2 - \eta)}{\sinh (\mu_1 - \mu_2)} \ \cosh (2 \xi ^+ - \mu _1 + \mu _2 + 2 \eta ) \ \alpha (\mu _1)   \widehat{ \delta}  (\mu _2) \notag \\[1ex]
&- \frac{\sinh (2 \mu _2)}{\sinh (2 \mu _2 + \eta)} \ \frac{\sinh (\mu _2 + \mu _3) \sinh (\mu _2 - \mu _3 - \eta )}{\sinh (\mu _2 - \mu _3) \sinh (\mu _2 + \mu _3 + \eta )} \ \frac{\sinh (\mu _1 - \mu_3 + \eta) \sinh (\mu _1 + \mu_3 + 2 \eta)}{\sinh (\mu _1 - \mu _3) \sinh (\mu _1 + \mu_3 + \eta)} \ \times \notag \\[1ex]
&\times \ \frac{\sinh (\mu_2 - \mu_1 - \eta)}{\sinh (\mu_2 - \mu_1)} \  \cosh (2 \xi ^+ + \mu _1 - \mu _2 + 2 \eta ) \ \widehat{ \delta} (\mu _1) \alpha (\mu _2) + \frac{\sinh (\mu_1 + \mu_2 + 2 \eta)}{\sinh (\mu_1 + \mu_2 + \eta)} \ \times \notag \\[1ex]
&\left. \times \prod _{i=1}^2  \frac{\sinh (\mu _i - \mu_3 + \eta) \sinh (\mu _i + \mu_3 + 2 \eta)}{\sinh (\mu _i - \mu _3) \sinh (\mu _i + \mu_3 + \eta)} \ \cosh (2 \xi ^+ + \mu _1 + \mu _2 + 3 \eta ) \ \widehat{ \delta} (\mu _1) \widehat{ \delta} (\mu _2) \right) ,
\end{align}
and
\begin{align}
\label{3b3-tilde}
&\widetilde{b} ^{(3)}_3 (\mu_1 ,  \mu_2 ,  \mu_3) = \left( - \frac{\psi^+}{\kappa ^+} \right)^3 \left( \prod _{i=1}^3 \frac{\sinh (2 \mu _i)}{\sinh (2 \mu _i + \eta)} \ \prod _{j >i}^3
\frac{\sinh (\mu _i + \mu_j)}{\sinh (\mu _i + \mu _j + \eta)} \cosh (3 \xi ^+ - \mu _1 - \mu _2  - \mu _3 + 3 \eta ) \ \times \right. \notag \\[1ex]
& \times \alpha (\mu _1) \alpha (\mu _2) \alpha (\mu _3) - \frac{\sinh (\mu _1 + \mu_2)}{\sinh (\mu _1 + \mu _2 + \eta)} \ \prod _{i=1}^2 \ \frac{\sinh (2 \mu _i)}{\sinh (2 \mu _i + \eta)} \ \frac{\sinh (\mu_i - \mu_3 - \eta)}{\sinh (\mu_i - \mu_3)} \ \times \notag \\[1ex]
& \times \cosh (3 \xi ^+ - \mu _1 - \mu _2  + \mu _3 + 4 \eta ) \ \alpha (\mu _1) \alpha (\mu _2) \widehat{ \delta}  (\mu _3) - \frac{\sinh (\mu _1 + \mu_3)}{\sinh (\mu _1 + \mu _3 + \eta)} \ 
\prod _{\substack{ i =1\\ i \neq 2}}^3 \ \frac{\sinh (2 \mu _i)}{\sinh (2 \mu _i + \eta)} \ \times \notag \\[1ex]
& \times \frac{\sinh (\mu_i - \mu_2 - \eta)}{\sinh (\mu_i - \mu_2)} \ 
\cosh (3 \xi ^+ - \mu _1 + \mu _2  - \mu _3 + 4 \eta ) \ \alpha (\mu _1) \alpha (\mu _3) \widehat{ \delta}  (\mu _2) - \frac{\sinh (\mu _2 + \mu_3)}{\sinh (\mu _2 + \mu _3 + \eta)} \ \times \notag \\[1ex]
& \prod _{i=2}^3 \ \frac{\sinh (2 \mu _i)}{\sinh (2 \mu _i + \eta)} \ \frac{\sinh (\mu_i - \mu_1 - \eta)}{\sinh (\mu_i - \mu_1)}  \ \cosh (3 \xi ^+ + \mu _1 - \mu _2  - \mu _3 + 4 \eta ) \ \alpha (\mu _2) \alpha (\mu _3) \widehat{ \delta}  (\mu _1) \notag \\[1ex]
&+ \frac{\sinh (2 \mu _1)}{\sinh (2 \mu _1 + \eta)} \  \prod _{i=2}^3 \frac{\sinh (\mu_1 - \mu_i - \eta)}{\sinh (\mu_1 - \mu_i)} \ \frac{\sinh (\mu_2  + \mu_3 + 2 \eta )}{\sinh (\mu_2  + \mu _3 + \eta)} \ \cosh (3 \xi ^+ - \mu _1 + \mu _2  + \mu _3 + 5 \eta ) \times  \notag \\[1ex]
&\times  \alpha (\mu _1) \widehat{ \delta} (\mu _2) \widehat{ \delta} (\mu _3) \ + \frac{\sinh (2 \mu _2)}{\sinh (2 \mu _2 + \eta)} \ \prod _{\substack{ i =1\\ i \neq 2}}^3 \  \frac{\sinh (\mu_2 - \mu_i - \eta)}{\sinh (\mu_2 - \mu_i)} \ \frac{\sinh (\mu_1  + \mu_3 + 2 \eta )}{\sinh (\mu_1  + \mu _3 + \eta)} \times  
\notag \\[1ex]
&\times \cosh (3 \xi ^+ + \mu _1 - \mu _2  + \mu _3 + 5 \eta ) \  
\alpha (\mu _2) \widehat{ \delta} (\mu _1) \widehat{ \delta} (\mu _3) \
+ \frac{\sinh (2 \mu _3)}{\sinh (2 \mu _3 + \eta)} \  \prod _{i=1}^2 \frac{\sinh (\mu_3 - \mu_i - \eta)}{\sinh (\mu_3 - \mu_i)} \times  \notag \\[1ex]
& \times \ \frac{\sinh (\mu_1  + \mu_2 + 2 \eta )}{\sinh (\mu_1  + \mu _2 + \eta)} \ \cosh (3 \xi ^+ + \mu _1 + \mu _2  - \mu _3 + 5 \eta ) \ \alpha (\mu _3) \widehat{ \delta} (\mu _1) \widehat{ \delta} (\mu _2) \notag \\[1ex]
&\left. -  \prod _{i=1}^3 \ \prod _{j >i}^3 \ \frac{\sinh (\mu_i + \mu_j + 2 \eta)}{\sinh (\mu_i + \mu_j + \eta)} \   \cosh (3 \xi ^+ + \mu _1 + \mu _2  + \mu _3 + 6 \eta ) \ \widehat{ \delta}  (\mu _1) \widehat{ \delta} (\mu _2) \widehat{ \delta} (\mu _3) \right) .
\end{align}
The action of $t(\lambda)$ \eqref{transfer-matrix} on $\widetilde{\Psi} _3 ( \mu_1 ,  \mu_2 ,  \mu_3)$, obtained by a straightforward calculations using evident generalization of the formulas \eqref{ABBOmega}, \eqref{DBBOmega} and  \eqref{CBBOmega} and subsequent rearranging of terms, is give by
\begin{equation}
\label{t-on-Psi3-til}
\begin{split} 
t(\lambda) \widetilde{\Psi} _3 ( \mu_1 ,  \mu_2 ,  \mu_3) &= \Lambda _3 (\lambda , \{\mu _i \}) \widetilde{\Psi} _3 ( \mu_1 ,  \mu_2 ,  \mu_3) + \sum _{i=1}^3 \frac{ \sinh (\eta) \sinh (2 (\lambda + \eta ))}{\sinh (\lambda - \mu _i) \sinh (\lambda + \mu _i + \eta )} \times \\[1ex]
&\times \ \kappa ^+ \sinh (\xi^+ - \mu _i ) \ F_3 (\mu _i ; \{\mu _j \} _{j \neq i}) \ \widetilde{\Psi}_3 ( \lambda ,  \{\mu _j \} _{j \neq i}) ,
\end{split}
\end{equation}
where the eigenvalue $\Lambda _3 (\lambda , \{\mu _i \})$ is given in \eqref{Lambda3} and the function $F _3 (\mu _i ; \{\mu _j \} _{j \neq i})$ in \eqref{BE-3}. 

With the aim of adding some extra terms, multiplied by some arbitrary coefficients and in this sense generalizing $\widetilde{\Psi} _3 ( \mu_1 ,  \mu_2 ,  \mu_3)$ in such a way that the action of $t(\lambda)$ \eqref{t-on-Psi3-til} is preserved, we observe the following six identities. The first three identities, which are straightforward generalization of the identities \eqref{identity-Psi2-L1} and \eqref{identity-Psi2-L2} relevant in the $M=2$ case, are given by
\begin{equation}
\label{identity-Psi3-Li}
\Lambda _{2} (\lambda, \{ \mu _j \}_{j\neq i}^3) - \Lambda _{3} (\lambda, \{ \mu _j \}_{j=1}^3)
= \kappa ^+ \sinh ( \xi^+ - \lambda ) \frac{\sinh (\eta) \sinh ( 2 (\lambda + \eta) )}{\sinh (\lambda - \mu _i) \sinh(\lambda + \mu _i + \eta)} F_3 (\lambda ; \{\mu _j \}_{j\neq i}^3) ,
\end{equation}
here $i = 1, 2, 3$ and the other three identities, which are generalization of the identity \eqref{identity-Psi2-F1,2}  in the $M=2$ case, are 
\begin{align}
\label{identity-Psi3-Fi,j}
&\frac{F_3 (\mu _j ; \{\mu _k \}_{k\neq j}^3) \ F_2  ( \mu _i; \{\mu _k \}_{k\neq i, j}^3) - F_3 (\mu _i ; \{\mu _k \}_{k\neq i}^3) \ F_3 (\mu _j ; \lambda , \{\mu _k \}_{k\neq i, j}^3 )}{\sinh (\lambda - \mu _i) \sinh(\lambda + \mu _i + \eta)} + \notag \\[1ex]
+ &\frac{F_3 (\mu _i ; \{\mu _k \}_{k\neq i}^3) \ F_2  ( \mu _j; \{\mu _k \}_{k\neq i, j}^3) - F_3 (\mu _j ; \{\mu _k \}_{k\neq j}^3) \ F_3 (\mu _i ; \lambda , \{\mu _k \}_{k\neq i, j}^3)}{\sinh (\lambda - \mu _j) \sinh(\lambda + \mu _j + \eta)} = 0 , 
\end{align}
here $i < j$, $\ i = 1, 2$, and $j=2,3$.  Therefore the general form of the Bethe vector\break  $\widetilde{\Psi }_3 ( \mu_1 ,  \mu_2, \mu_3, C_1, C_2, C_3 )$ is given by
\begin{equation}
\label{Psi3-tilde}
\widetilde{\Psi }_3 ( \mu_1 ,  \mu_2, \mu_3, C_1, C_2, C_3 ) = \widetilde{\Psi }_3 ( \mu_1 ,  \mu_2, \mu_3 ) + C_3 \ \frac{\psi^+}{\kappa ^+}  \ \sum _{i=1}^3 \sinh (\xi ^+ - \mu _i) F_3 (\mu _i ; \{\mu _j \}_{j \neq i}) \widetilde{\Psi }_2( \{\mu _j \}_{j \neq i} , C_1, C_2)  ,
\end{equation}
where $C_3$ does not depend on $ \{\mu _i \} _{i=1}^3$ and $\widetilde{\Psi }_2(  \lambda ,  \mu_i , C_1, C_2)$ is given in \eqref{Psi2-tilde}. Due to \eqref{t-on-Psi3-til} and the above identities \eqref{identity-Psi3-Li} -- \eqref{identity-Psi3-Fi,j} it is straightforward to check that the off-shell action of transfer matrix $t(\lambda)$ on $\widetilde{\Psi }_3 ( \mu_1 ,  \mu_2, \mu_3, C_1, C_2, C_3 )$ is 
\begin{equation}
\label{t-on-Psi3-tilde}
\begin{split} 
t(\lambda) \widetilde{\Psi }_3 ( \mu_1 ,  \mu_2, \mu_3, C_1, C_2, C_3) &= \Lambda _3 (\lambda , \{\mu _i \}) \Psi _3 ( \mu_1 ,  \mu_2 ,  \mu_3, C_1, C_2, C_3 ) + \sum _{i=1}^3 \frac{ \sinh (\eta) \sinh (2 (\lambda + \eta ))}{\sinh (\lambda - \mu _i) \sinh (\lambda + \mu _i + \eta )} \times \\[1ex]
&\times \ \kappa ^+ \sinh (\xi^+ - \mu _i ) \ F_3 (\mu _i ; \{\mu _j \} _{j \neq i}) \ \Psi _3 ( \lambda ,  \{\mu _j \} _{j \neq i}, C_1, C_2, C_3 ) .
\end{split}
\end{equation}
By setting $C _1 = \displaystyle{\frac{1-2 e^{2\eta} -2 e^{4\eta} - e^{6\eta}}{1- e^{6\eta}}}, C _2 = - \tanh (\eta)$ and $C _3 =1$ in \eqref{Psi3-tilde} we obtain the corresponding Bethe vector $\Psi _3 ( \mu_1 ,  \mu_2 ,  \mu_3)$ \eqref{Psi3}, i.e.
\begin{equation}
\label{Psi3-ours}
\Psi _3 ( \mu_1 ,  \mu_2 ,  \mu_3) = \widetilde{\Psi}_3( \mu_1 ,  \mu_2 ,  \mu_3, C _1 = \frac{1-2 e^{2\eta} -2 e^{4\eta} - e^{6\eta}}{1- e^{6\eta}}, C _2 = - \tanh (\eta), C _3 = 1).
\end{equation}

Although it would be natural to continue this approach and present here the Bethe vector $\widetilde{\Psi }_4 ( \mu_1 ,  \mu_2, \mu_3, \mu_4 )$,  which in the scaling limit corresponds to the Bethe vector of the XXX chain \cite{CAMS}, it turns out that the expressions for the coefficients functions $\widetilde{b} ^{(i)}_4 (\mu_1, \dots , \mu_i ; \mu_{i+1},  \ldots , \mu_4)$ are cumbersome, not admitting any compact form. For this reason we have decided not present them here. 

Indeed, the main obstacle in this approach is the lack of the closed form for the coefficients functions $\widetilde{b} ^{(i)}_M (\mu_1, \ldots , \mu_i ; \mu_{i+1},  \ldots , \mu_M)$ of the Bethe vector $\widetilde{\Psi }_M ( \mu_1 ,  \ldots, \mu_M )$, whose scaling limit corresponds to the Bethe vector of the XXX chain, for an arbitrary natural number $M$. All the necessary identities are know, the $M$ identities of the first type
\begin{equation}
\label{identity-PsiM-Li}
\Lambda _{M-1} (\lambda, \{ \mu _j \}_{j\neq i}^M) - \Lambda _{M} (\lambda, \{ \mu _j \}_{j=1}^M)
= \kappa ^+ \sinh ( \xi^+ - \lambda ) \frac{\sinh (\eta) \sinh ( 2 (\lambda + \eta) )}{\sinh (\lambda - \mu _i) \sinh(\lambda + \mu _i + \eta)} F_M (\lambda ; \{\mu _j \}_{j\neq i}^M) ,
\end{equation}
here $i = 1, \ldots, M$  and the $\frac{M(M-1)}{2}$ identities of the second type
\begin{align}
\label{identity-PsiM-Fi,j}
&\frac{F_M (\mu _j ; \{\mu _k \}_{k\neq j}^M) \ F_{M-1}  ( \mu _i; \{\mu _k \}_{k\neq i, j}^M) - F_M (\mu _i ; \{\mu _k \}_{k\neq i}^M) \ F_M (\mu _j ; \lambda , \{\mu _k \}_{k\neq i, j}^M )}{\sinh (\lambda - \mu _i) \sinh(\lambda + \mu _i + \eta)} + \notag \\[1ex]
+ &\frac{F_M (\mu _i ; \{\mu _k \}_{k\neq i}^M) \ F_{M-1}  ( \mu _j; \{\mu _k \}_{k\neq i, j}^M) - F_M (\mu _j ; \{\mu _k \}_{k\neq j}^M) \ F_M (\mu _i ; \lambda , \{\mu _k \}_{k\neq i, j}^M)}{\sinh (\lambda - \mu _j) \sinh(\lambda + \mu _j + \eta)} = 0 , 
\end{align}
here $i < j$, $\ i = 1, 2, \ldots , M-1$, and $j=2,3, \ldots , M$. The most general form of the Bethe vector, for an arbitrary positive integer $M$, is given as a sum of a particular vector and a linear combination of lower order Bethe vectors that correspond to the same eigenvalue
\begin{equation}
\label{general-Psi}
\begin{split}
\widetilde{\Psi }_M ( \{\mu_i \}_{i=1}^M ,  \{ C_j \}_{j=1}^M ) &= \widetilde{\Psi }_M ( \mu_1 ,  \ldots, \mu_M ) + C_M \ \frac{\psi^+}{\kappa ^+}  \ \sum _{i=1}^M \sinh (\xi ^+ - \mu _i) F_M (\mu _i ; \{\mu _j \}_{j \neq i}^M) \times \\
&\times \widetilde{\Psi }_{M-1}( \{\mu_j \}_{j\neq i}^M ,  \{ C_k \}_{k=1}^{M-1} ) .
\end{split}
\end{equation}

Unfortunately, this approach cannot be used in general case due to the lack of the closed form for the coefficients functions of the Bethe vector $\widetilde{\Psi }_M ( \mu_1 ,  \ldots, \mu_M )$. On the other hand, as it is evident form the formulae \eqref{bM-2} --  \eqref{bM-M}, the recurrence procedure we propose is clearly advantages providing basically the same formulae, up to the multiplicative factors, like in the case of the XXX Heisenberg spin chain \cite{CAMS}, for the coefficients functions 
$b ^{(i)}_M (\mu_1, \dots , \mu_i ; \mu_{i+1},  \ldots , \mu_M)$ of the Bethe vector $\Psi _M ( \mu_1 ,  \ldots , \mu_M )$, besides $b ^{(1)}_M (\mu_1 ; \mu_{2},  \ldots , \mu_M)$ which is given explicitly in \eqref{bM-1}.

\end{document}